\documentclass[conference]{IEEEtran}
\usepackage{cite}
\usepackage[pdftex]{graphicx}
\usepackage{amsmath}
\usepackage{algorithmic}
\usepackage{array}
\usepackage{xcolor}
\definecolor{darkblue}{RGB}{0, 51, 153}
\definecolor{lightblue}{RGB}{245, 245, 255}
\colorlet{lightblue}{lightblue!30!white}

\usepackage{multirow}
\usepackage{hyperref}
\usepackage{booktabs}
\usepackage{comment}

\usepackage{tabularx}

\ifCLASSOPTIONcompsoc
 \usepackage[caption=false,font=normalsize,labelfont=sf,textfont=sf]{subfig}
\else
 \usepackage[caption=false,font=footnotesize]{subfig}
\fi
\usepackage{fixltx2e}
\usepackage{stfloats}
\usepackage{url}
\usepackage[normalem]{ulem}
\usepackage{tcolorbox}
\usepackage{enumitem}

\newtcolorbox{remarkbox}[1][]{
    colback=white,
    colframe=gray,
    fonttitle=\bfseries,
    title=Remark,
    arc=1pt,
    before skip=5pt, 
    after skip=-1pt,  
    boxsep=3pt,   
    left=4pt,     
    right=4pt,    
    top=2pt,      
    bottom=2pt,   
    #1
}

\newlist{rqs}{enumerate}{1}
\setlist[rqs]{label*=\textbf{RQ\arabic*}}

\newcommand{\paragraphY}[1]{\vskip 3pt \noindent \emph{#1} \hskip .05in}

\newcommand{\buildmode}{2}

\ifnum\buildmode=0
  \newenvironment{camready}{\color{blue}}{}
  \newenvironment{arxiv}{\color{red}}{}
\fi
\ifnum\buildmode=1
  \newenvironment{camready}{}{}
  \excludecomment{arxiv}
\fi
\ifnum\buildmode=2
  \excludecomment{camready}
  \newenvironment{arxiv}{}{}
\fi

\begin{document}

\title{``What I See is What I Hear'': Deepfake Detection Across Diverse Hearing Abilities}

\author{%
\IEEEauthorblockN{%
    Magdalena Pasternak\IEEEauthorrefmark{1},
    Malvika Jadhav\IEEEauthorrefmark{1},
    Palavi V. Bhole\IEEEauthorrefmark{2},
    Aviva Smith\IEEEauthorrefmark{1},
    Elaina Trapatsos\IEEEauthorrefmark{2},\\
    Vincent Bindschaedler\IEEEauthorrefmark{1},
    Roshan Peiris\IEEEauthorrefmark{2},
    Ersin Uzun\IEEEauthorrefmark{2},
    Patrick Traynor\IEEEauthorrefmark{1},
    Matthew Wright\IEEEauthorrefmark{2},
    Kevin R.\ B.\ Butler\IEEEauthorrefmark{1}}
    \IEEEauthorblockA{\IEEEauthorrefmark{1}University of Florida\\
    \{mpasternak, jadhav.m, asmith37, vbindschaedler, traynor, butler\}@ufl.edu}
    \IEEEauthorblockA{\IEEEauthorrefmark{2}Rochester Institute of Technology\\
    \{pb7169, ejt7508, roshan.peiris, ersin.uzun, matthew.wright\}@rit.edu}
}

\IEEEoverridecommandlockouts
\makeatletter\def\@IEEEpubidpullup{6.5\baselineskip}\makeatother
\IEEEpubid{\parbox{\columnwidth}{
		Network and Distributed System Security (NDSS) Symposium 2027\\
		22--26 March 2027, Seoul, Republic of Korea\\
		ISBN 978-1-970672-09-1\\  
		https://dx.doi.org/10.14722/ndss.2027.230307\\
		www.ndss-symposium.org
}
\hspace{\columnsep}\makebox[\columnwidth]{}}

\maketitle

\begin{abstract}
The proliferation of audiovisual deepfakes has lowered the cost of fraud, impersonation, and misinformation, but their success ultimately depends on human perception. Detection requires integrating auditory and visual cues, yet security and privacy research has largely overlooked d/Deaf and hard-of-hearing (DHH) populations. We address this gap with an in-person, mixed-methods study of 80 participants: 31 hearing persons (HPs), 15 hard-of-hearing (HoH) participants, 17 d/Deaf participants, and 17 cochlear implant (CI) users. Each participant judged the authenticity of 30 clips, where manipulations spanned text-to-speech, voice conversion, lip-sync, or face-swap. DHH participants were less accurate than HPs overall (76.4\% vs.\ 88.0\%, p\textless{}.001), primarily because they more often classified authentic clips as manipulated (FPR: 29.7\% vs.\ 11.2\%). Differences depended strongly on the manipulated channel. For audio-only manipulations, HoH participants matched HPs (90.0\% vs.\ 90.3\%), followed by CI users (79.4\%) and d/Deaf participants (41.2\%). When clips contained an audiovisual manipulation, accuracy clustered between 84\% and 87\%, although performance still varied by manipulation method. Our work systematically characterizes how deepfakes affect DHH populations, highlighting the asymmetric risks audiovisual manipulations may pose to groups with different hearing abilities and the need for accessible, tailored defenses that support all users.
\end{abstract}

%
\IEEEpeerreviewmaketitle

\section{Introduction}\label{sec:intoduction}
Advances in generative AI have lowered the cost and expertise required to produce, personalize, and distribute synthetic media. Deepfakes can convincingly imitate reality by exploiting the limits of human perception~\cite{cooke2025cointoss}.They can facilitate financial fraud~\cite{pump_and_dump, Magramo_2024, Allen_2025} and extortion~\cite{romance_scam}, erode public trust~\cite{social_impact_deepfakes}, and polarize communities~\cite{deepfas_polarize_politics, df_threat_marginalized}. As generation tools become cheaper and easier to use~\cite{chesney2019deep}, deepfake attacks can be produced and distributed globally, making them an increasingly pervasive digital threat.

Current technical countermeasures offer limited protection. Automated detectors remain constrained by dataset coverage and generalization~\cite{9721302, groh2022deepfake, layton2024sok, generalize}, while provenance standards such as C2PA depend on widespread adoption throughout the media life-cycle~\cite{C2PA_specs}. Fact-checking services~\cite{Fact_Check} and community moderation can corroborate claims~\cite {grain_of_salt}, but their coverage is selective, and their judgments often arrive after initial exposure. The burden of identifying synthetic media thus falls on viewers who must protect themselves from deception based on the evidence they can access. 

While prior work has studied human deepfake detection~\cite{zhou2021joint, hashmi2024unmaskingillusionsunderstandinghuman, Fatima2026CHI}, it has largely assumed a default viewer and overlooked how hearing ability shapes susceptibility. This gap is consequential, as more than 430 million people worldwide have disabling hearing loss, and approximately $15\%$ (37.5 million) US adults report some difficulty hearing~\cite{worldhealthorganization_2025_deafness, nidcd_hearing_stat}. 

Hearing loss varies widely, as do identity and communication practices within d/Deaf and Hard-of-Hearing (DHH) communities. Hard-of-Hearing (HoH) generally refers to people with mild-to-severe hearing loss who often communicate through spoken language. Lowercase `deaf' refers to a medical condition of significant hearing loss, whereas `Deaf' denotes a cultural and social identity commonly associated with sign language, and d/Deaf encompasses both. Some DHH individuals use cochlear implants (CIs), with more than one million devices implanted worldwide by 2022~\cite{zeng_2022_milion_ci}. Unlike hearing aids, which amplify sound, CIs provide auditory access by electrically stimulating the auditory nerve~\cite{nid_ci_ha}. 

These differences in hearing loss, identity, and communication practice shape how audiovisual content is encountered and which evidence is accessible. For DHH users, access to audiovisual content accordingly depends on combinations of residual hearing~\cite{stevenson2017multisensory}, speech-reading~\cite{bernstein2022lipreading}, assistive technologies such as hearing aids or CIs, and accessibility infrastructure such as Video Relay Services~\cite{fcc_vrs}, captions~\cite{nid_captioning}, and sign-language interpretation. While these access layers support content availability, they can also variously preserve, omit, or add information about speaker identity, affect, uncertainty, non-speech sounds, and context. Audiovisual deepfakes therefore create a layered authenticity problem involving both the source content and the accessible representation through which users encounter it.

The small body of work that looked beyond default user profiles examined deception in audio channels. For blind and low-vision users, these studies examined the channel they rely on~\cite{sharevski2024blind, han2024uncoveringblind}. In contrast, for DHH individuals, specifically CI users, the only prior study examines a channel to which access is technologically mediated~\cite{pasternak2025ci}. This further highlights the neglect of accessibility in usable security~\cite{renaud2022accessible}. 

We address this gap by systematically quantifying how hearing profile shapes susceptibility across audiovisual deepfakes. Through an in-person, mixed-methods study with 80 participants, including 31 hearing persons (HPs), 15 hard-of-hearing (HoH), 17 d/Deaf participants, and 17 CI users, we compare detection accuracy across manipulation types and characterize DHH users' perceptions and detection strategies. We ask the following research questions:

\begin{rqs}[leftmargin=3em]
 \item How does audiovisual deepfake detection accuracy differ between HP and DHH participants and across DHH subgroups (HoH, d/Deaf, CI users)?\label{rq:q1}
 \item How does detection accuracy differ across visual-only, audio-only, and combined audiovisual manipulations?\label{rq:q2}
 \item Which perceptual cues and heuristics do HP and DHH participants report when explaining their decisions, and how are they associated with correct classifications?\label{rq:q3}
\end{rqs}

\noindent We thus make the following contributions:
\begin{itemize}[leftmargin=1.8em]
 \item \textbf{An evaluation of audiovisual deepfake detection across hearing profiles.} We conduct an in-person, mixed-methods study with 80 participants. DHH participants reached $76.4\%$ overall accuracy (subgroups: $73.1\%$--$78.6\%$), compared with $88.0\%$ for HPs, a gap driven primarily by systematic rejection of authentic clips (TNR $70.3\%$ vs.\ $88.8\%$). We further characterize confidence, calibration, and error patterns. 
 \item \textbf{Comparative analysis across manipulation types.} We characterize how sensory access and manipulation type shape susceptibility. Audio-only detection depended on auditory access: HoH participants matched HPs ($90.0\%$ vs.\ $90.3\%$), while d/Deaf participants fell to $41.2\%$, indistinguishable from their response to real clips. Manipulation types also produced distinct subgroup patterns; CI users, for example, achieved $90.8\%$ on face-swaps but $77.5\%$ on lip-syncs. 
 \item \textbf{Qualitative analysis of perceptual detection strategies.} We double-coded participants' explanations and analyzed how perceptual cues and reasoning strategies varied across hearing profiles. Cue use reflected sensory access and attention: d/Deaf participants focused on the mouth region, whereas CI users attended more closely to facial geometry and shape.
\end{itemize}

Our findings show that authenticity judgments depend on the evidence viewers can access. HPs and DHH subgroups exhibited distinct strengths, error patterns, and cue-use profiles, yet common guidance directs all viewers to inspect the same features, such as lip movements or blinking~\cite{mit_medialit_deepfakes}, and defenses often take a one-size-fits-all approach. Under unequal access, deepfake deception is both a \emph{security} problem, because attackers choose the manipulation, and a \emph{usable-security} problem, because the target's access profile determines which evidence users can inspect and protections may rely on cues some users cannot examine. Detection tools and policies must therefore be designed and evaluated across access profiles, rather than built for a default user and retrofitted after deployment.

\section{Background}\label{sec:background}
\subsection{Deepfake Generation}\label{back_df_gen}
Advances in generative media now enable adversaries to produce realistic, manipulated, and synthetic content cheaply, at scale, and with minimal technical skill~\cite{chesney2019deep, ho2022video}. These manipulations, known as deepfakes, may employ various techniques and span both the audio and visual channels of media consumption~\cite{mubarak2023survey}. Audio deepfakes are primarily produced using text-to-speech (TTS) and voice conversion (VC) techniques~\cite{almutairi2022review}. TTS synthesizes new speech utterances directly from text while mimicking a target's voice. In contrast, VC modifies an existing sample to match the target's voice characteristics while preserving the underlying linguistic content and temporal structure~\cite{khanjani2023audio}. Visual deepfakes~\cite{mubarak2023survey} include \textit{face-swap}, which replaces the original face with another while aligning facial landmarks and preserving realistic pose and expression~\cite{perov2020deepfacelab}, and \textit{lip-sync}, which alters mouth movements to match a new audio track~\cite{prajwal2020lip}, \textit{reenactment} which transfers facial movements or expressions across identities~\cite{thies2016face2face}, and \textit{facial manipulation}, which alters attributes or expressions within existing media~\cite{mubarak2023survey}. Audiovisual deepfakes combine these techniques into attacks that target the visual channel, the audio channel, or their alignment. Recent generative techniques synthesize entire clips simultaneously even from text prompts alone~\cite{align_latens}.

\subsection{Automated Detection}\label{back_df_auto_detect}
To counter these threats, a large body of research has focused on developing automated deepfake detectors~\cite{todisco2019asvspoof,FakeAVCeleb, zhou2021joint}. Unfortunately, their strong performance on benchmark datasets does not necessarily translate into robustness in deployment. Recent work shows that detector performance degrades under distribution shift, unseen manipulation methods, post-processing, and adversarial adaptation~\cite{wang2024deepfake, generalize}, and that benchmark datasets can encode unrealistic artifacts or imbalances that might inflate reported accuracy~\cite{layton2024sok}. Detection also inherits this asymmetry because the artifacts it looks for are the residue of altering an authentic recording, including mismatches between a modified channel and an untouched one~\cite{zhou2021joint}, and fabrication leaves no such residue. Even if it were universally adopted, automated detection would remain an incomplete defense layer, and humans would remain the final line of defense against manipulated and fabricated content.

\subsection{Human Detection}\label{back_hum_detect}
Ultimately, deepfakes succeed only when they deceive people and a viewer accepts manipulated content as authentic, forms beliefs based on it, or acts on it. Consequently, deepfakes are not only a media forensics problem but also a human-centered security challenge. Whether an attack succeeds depends on the cues viewers perceive, the trust they place in the source, and how they reason under uncertainty~\cite{muller2022human, warren2024better, cooke2025cointoss}. 

This perspective has motivated a growing body of work on human deepfake detection, investigating how people distinguish authentic from manipulated media~\cite{cooke2025cointoss, warren2024better, muller2022human, groh2022deepfake, groh2024human, hashmi2024unmaskingillusionsunderstandinghuman, korshunov2020deepfake, Fatima2026CHI}. These studies show that detection strategies draw on visual and auditory cues, cognitive heuristics, and contextual knowledge, and that detection performance varies across datasets, manipulation techniques, and presentation modalities~\cite{human_performance_review}. Unlike prompted study settings, real-life consumption provides no warning, and a broader context of platform, apparent source, surrounding claims, and requested action can shape whether it is believed~\cite{ruffin2024does, mink2022deepphish, lovato_diverse_2024}. 

Security consequences extend beyond unsafe trust in manipulated content. Authentic evidence may also be dismissed as fabricated, allowing those it implicates to avoid accountability by falsely claiming that damaging true information is fake, a dynamic known as the liar's dividend~\cite{schiff2025liar}. Additionally, warnings about manipulated content might further increase the likelihood of judging an authentic clip as fake, without improving the ability to spot deepfakes~\cite{ternovski2022negative}.

\subsection{DHH Users}\label{back_dhh_users}
Whether a viewer accepts or rejects a clip depends on what evidence the viewer can access. Populations with systematically different sensory access face distinct risk surfaces~\cite{pasternak2025ci}. Human detection studies rarely test this and recruit broad participant pools that assume uniform access to audiovisual information, with few exceptions addressing audio deception of blind and low vision individuals~\cite{sharevski2024blind, han2024uncoveringblind} and CI users~\cite{pasternak2025ci}. 

Prior security and privacy research on DHH users shows that accessibility gaps lead to concrete security and privacy vulnerabilities, not just reduced usability~\cite{andrew2020review, tran2026toward, buckmann2025more}. For DHH users facing potential deepfakes, the information available to evaluate audiovisual content depends not only on auditory access but also on language background, literacy, communication practices, and the accessibility mechanisms through which the content is delivered. 

\subsubsection{Hearing and Assistive Devices}
DHH users differ substantially in the extent and form of auditory information available to them. HoH individuals may receive a degraded auditory signal, whereas d/Deaf users may receive little or no auditory input without an assistive device, and CI users receive a reconstructed one. Hearing aids primarily amplify residual acoustic information, whereas cochlear implants (CIs) bypass damaged inner-ear structures and directly stimulate the auditory nerve to provide auditory access~\cite{nid_ci_ha}. Although CIs broaden access to speech, they can also alter auditory perception, including pitch and prosodic information that are relevant to audio deepfake detection~\cite{pasternak2025ci}. DHH users may additionally supplement auditory information with sound manipulation, sound visualization, and speech-to-text~\cite{ohshiro2022creative}, or combine these with lip reading and contextual cues~\cite{desai2023understandingdhh}. 

\subsubsection{Signed Language and Literacy}
Many DHH individuals use signed languages, such as American Sign Language (ASL) in the United States~\cite{nih_asl}. Signed languages are not signed versions of spoken languages, but rather distinct natural languages with their own grammatical and semantic structures~\cite{nih_asl, bragg2019sign}. DHH users vary in the languages and modalities through which they primarily communicate and access information, with some using a signed language, others relying on spoken or written language, and many using a~combination~\cite {zazove2013deaf}. Literacy levels also vary substantially, with deaf readers showing lower average English reading proficiency than their hearing peers~\cite{zazove2013deaf}. These differences shape how written information is accessed and interpreted, requiring security-relevant text to be designed to accommodate a wide range of literacy levels~\cite{sharevski_dhh}.

\subsubsection{Mediated Access}
Captions provide text-based access to auditory content and serve as an important accessibility mechanism across educational and social platforms~\cite{nid_captioning, mack2020socialapp}. Automated captions prioritize spoken linguistic content and can introduce recognition errors~\cite{kuhn2024asr, alonzo2022beyondsubtitles}, whereas manual captions can incorporate contextual information such as speaker identification and descriptions of non-speech sounds~\cite{nid_captioning, may2023nonspeech}. Captions therefore selectively preserve and transform information relative to the original audiovisual content rather than providing an equivalent representation~\cite {mcdonnell_captions}. 

Captions change which evidence remains available for judging its authenticity; transcript-only speech drops detection near chance, improving when audio is added~\cite{groh2024human}. Captions may preserve linguistic content while omitting acoustic artifacts that reveal manipulation and compete with other visual information for attention~\cite{kushalnagar2014classroom}, which deepfake detection requires. 

For signed-language users, interpretation similarly mediates access, preserving some information while transforming or omitting other features. Unlike most captioning, interpretation is interactive and extends beyond transferring linguistic content~\cite{bragg2019sign}. Interpreters' choices about emphasis, affect, and context shape how a message is received~\cite{haug2017deafleaders}, and can also shape how others perceive Deaf sign-language users, with implications for identity and agency captured by Young et al.'s concept of the \textit{``translated deaf self''}~\cite{young2020translated}.

\section{Threat Model}\label{sec:threat_model}

We consider an adversary who uses publicly available tools to generate and deliver a short, recorded talking-head video that manipulates an individual's identity, speech, or both. The adversary selects an impersonated subject and message, obtains the required reference material, generates and filters candidate outputs, and delivers the selected clip through a~channel the target uses. The target then judges its authenticity. Our study evaluates this judgment under controlled conditions rather than the consequential behavior that may follow it.

\paragraphY{Recipients and Access Profiles.}
Recipients differ in the evidence they can access in a clip. We represent this variation through an ``access profile'', which captures access to auditory and visual information, the reliability and salience of cues within each channel, communication modality, assistive technology, and the accessibility layer through which the content is received. Access profiles are not ordered by ability, as reduced access to one channel may increase reliance and scrutiny of another~\cite{bavelier2002cross}. They therefore describe which evidence a~recipient can access and use, rather than how much evidence the recipient can evaluate.

\paragraphY{Adversary Goals.} 
The adversary seeks either to deceive the target into accepting a manipulated clip as authentic (false negative) or to cause the target to reject authentic content as manipulated (false positive). False acceptance is the primary threat because it may lead the target to share the content, disclose information, transfer money, or follow fraudulent instructions. A secondary goal is rejection of authentic content, which erodes trust in authentic media.

False acceptance can arise through either \textit{perceptual deception} or \textit{evidentiary absence}. Perceptual deception occurs when diagnostic cues are within the target's sensory access but are misinterpreted or unnoticed. Evidentiary absence occurs when those cues lie in a channel that the target cannot reliably inspect. Therefore, the manipulation need not withstand scrutiny in that channel, only that its artifacts are inaccessible to the target. When the target relies on captions, transcription, or interpretation, the accessibility layer may convey the manipulated message while omitting cues needed to authenticate its source. This same layer, however, can also preserve or surface evidence of manipulation, as discussed in Section~\ref{disc:accessibility_success_attack}.

\paragraphY{Adversary Capabilities and Constraints.} 
We assume that the adversary can obtain source material (images, audio, or video) that captures the impersonated subject's biometric signature, that is, the facial appearance, voice, and speaking mannerisms that the manipulation forges. The adversary can also access generation tools and compute resources, and generate and filter multiple candidates to select the manipulated clip. They then deliver the clip through a channel the target uses. The availability of this material constrains which manipulations are feasible, because each manipulation forges a different component of the signature and requires a different sample. Speech synthesis conditioned on a short voice prompt can produce arbitrary new speech from a transcript, and voice conversion needs a short sample of the subject's speech to map an actor's delivery onto the subject's voice. For video, face-swapping can use a single image of the subject, while lip-sync reenactment needs minutes of talking-head video of that specific person (\S\ref{meth:stimuli}).

The adversary may use commercial services or self-hosted tools. Commercial services may require ownership verification for higher-fidelity cloning and impose provider-side logging, usage limits, content moderation, and account enforcement~\cite{ElevenLabs_safety, HeyGen_consent}, and self-hosted tools avoid these centralized controls. These capabilities reflect documented attacks, including voice impersonation in calls, face-swapping in fraudulent identity operations, and even during live calls~\cite{Quigley_2026, Intelligence_2026, Cross_2026_haotian, mask_drop_cought}. Section~\ref{meth:stimuli} explains how the selected generators represent manipulation capabilities observed in these attacks.

Finally, we assume that the adversary knows or can infer the target's population-level access profile from accessibility settings, habitual caption use, disclosed accommodations, participation in community spaces, or prior interactions. Based on this access profile, the adversary selects manipulations that maximize the chance of success at the minimal cost. For example, if the target is d/Deaf, then our findings indicate that the effort to get a good audio sample can be skipped, while the quality of mouth movements in the video should be checked carefully.

\paragraphY{Scope.}
We instantiate this threat model for hearing profiles by evaluating four manipulation families under audio-only, visual-only, and combined audiovisual conditions (\S\ref{meth:stimuli}) to determine whether their effectiveness varies across access profiles. We do not implement an optimized or individually targeted attack, and Section~\ref{disc:attack_channel} discusses what our results imply for adversarial tailoring and for access profiles beyond hearing. Our findings are limited to the tested families, implementations, and delivery conditions (\S\ref{disc:limitations}).

\section{Design and Methodology}\label{sec:methodology}

To characterize how audiovisual deepfake judgments vary across hearing profiles, we conducted an in-person mixed-methods study. Each participant evaluated 30 videos, resulting in 2,400 trial-level observations and 62.5 hours of interview data. The presented clips included real videos and audio-only, visual-only, and combined audiovisual manipulations.

\subsection{Participants}\label{meth:participants}

\subsubsection{Recruitment and Eligibility}
Over ten months, we recruited across all hearing groups through university mailing lists, campus fliers, direct in-person recruitment, personal networks, and community outreach. Eligible participants were at least 18 years old, proficient in English or American Sign Language (ASL), and able to attend an in-person session. We conducted all sessions in person. This format allowed us to verify inclusion criteria at enrollment, including CI use, which prior work found difficult to confirm remotely~\cite{pasternak2025ci}. It also elicited more substantive responses than online administration, where responses may be brief or fraudulent~\cite{warren2024better, liem2025botted}. We concluded recruitment for each DHH subgroup once their responses reached thematic saturation, with no new patterns emerging in participants' justifications. We assessed saturation on the coded cue corpus (\S\ref{sec:thematic}): per subgroup, we tracked the number of distinct cues that appeared in at least two participants and the number of additional participants required for the rarest such cues to appear across all manipulation classes. All participants who began a session completed the study, so no sessions were partial, and no trial outcomes were missing.

\subsubsection{Grouping and Classification} 
To assign participants into four hearing-profile groups, we used self-reported hearing identity, degree of hearing loss, and use of hearing assistive technology. We assigned participants who used at least one CI during the study to the CI group, regardless of whether they identified as d/Deaf or HoH. Among participants without CIs, assignment to the d/Deaf or HoH group followed their self-reported degree of hearing loss. We assigned participants with no hearing loss to the HP group (Table~\ref{table:hearing_groups}).

\subsubsection{Sample Composition}
Our cohort of 80 participants comprised 31 hearing persons (HPs) and 49 d/Deaf and Hard-of-Hearing (DHH) individuals (15 Hard-of-Hearing (HoH), 17 d/Deaf, and 17 cochlear implant (CI) users). Our 49-participant DHH cohort is among the largest in-person studies in security research with this at-risk population. 
\begin{camready}
We report the participants' demographics, prior exposure to deepfakes, and assumed deepfake prevalence in Appendix~\ref{app:demographic}, with further background analyses in the extended version~\cite{arxiv_version}.
\end{camready}
\begin{arxiv}
We report the participants' demographics, online media consumption, prior exposure to deepfakes, and perceptions of deepfake prevalence and concerns in Appendix~\ref{app:demographic}.
\end{arxiv}

\begin{table}[t!]
 \centering
 \caption{Criteria used to assign participants to their groups.}\vspace{-0.2cm}
 \label{table:hearing_groups}
 
 \begin{tabularx}{0.95\columnwidth}{@{} p{0.058\columnwidth} p{0.38\columnwidth} X @{}}
 \toprule
 \textbf{} &
 \textbf{Hearing Identity / Loss} &
 \textbf{Assistive technology} \\
 \midrule
 
 \textbf{HP} &
 no hearing loss&
 None. \\
 
 \addlinespace
 
 \textbf{HoH} &
 mild--moderate hearing loss&
 May use hearing aid(s); no CI \\
 
 \addlinespace
 
 \textbf{Deaf} &
 d/Deaf; profound hearing loss&
 May use hearing aid(s); no CI\\
 
 \addlinespace
 
 \textbf{CI} &
 HoH or d/Deaf; uses CI(s)&
 CI(s) + may use hearing aid(s)\\
 
 \bottomrule
 \end{tabularx}
\end{table}

\subsubsection{Accessibility Accommodations}
When scheduling, we asked whether participants required an ASL interpreter and arranged one when requested. Interpreters supported consent, instructions, and communication with the research team but did not interpret the stimuli. When interpreters became unavailable on short notice, we rescheduled affected sessions when possible. Otherwise, an ASL-proficient member of our research team conducted the session. As a final fallback, participants could communicate by text with a researcher present, and two participants ultimately used this option.

Finally, we did not provide captions during the detection task, as caption use and quality vary across participants and across manually written, automatically generated, and interpreted content. Captions also introduce a derived representation of the audio signal that can independently affect authenticity judgments~\cite{groh2024human}.
Including captioning without treating it as a separate experimental factor would thus make it difficult to distinguish responses associated with hearing profile from those associated with caption content and quality. We therefore presented the same unmediated audiovisual signal to all participants, providing a controlled baseline for comparing responses to the same audiovisual signal.

\begin{figure*}[!t]
 \centering
 \includegraphics[width=0.98\linewidth]{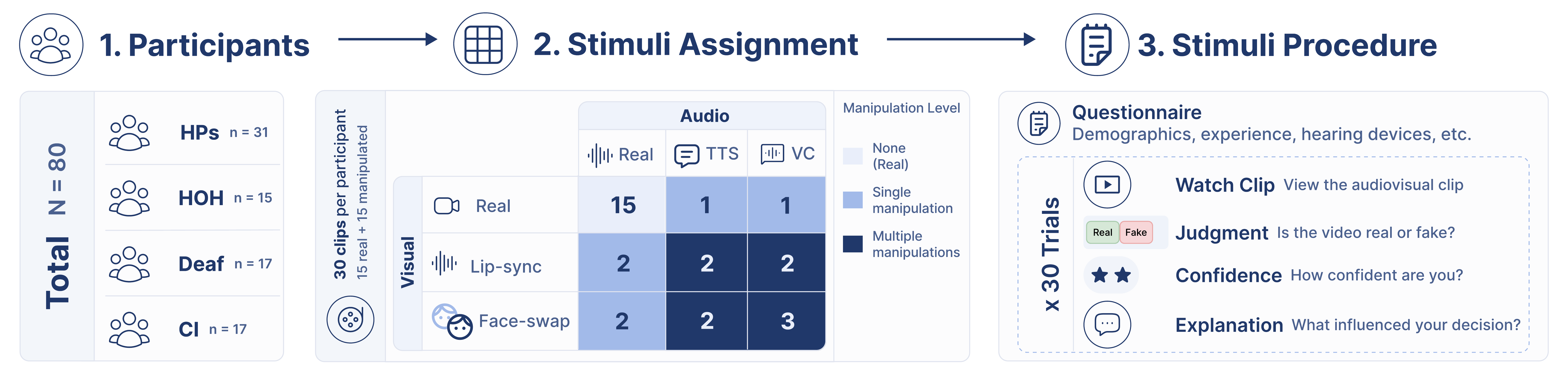}
 \caption{We conducted an in-person deepfake detection study with participants across 4 groups (HPs, HoH, d/Deaf, and CI users). Each participant viewed a randomized set of 30 clips to make a binary judgment (real/fake), reported confidence on a 1--5 Likert scale, and described the cues underlying their decision. After the video trials, participants completed a short post-study questionnaire.}
 \label{fig:study_procedure}
\end{figure*}

\subsection{Stimuli}\label{meth:stimuli} 
We focused on talking-head videos as facial appearance and voice are primary cues for recognizing and identifying a person. Deepfakes exploit these cues by depicting a target saying or doing something they did not~\cite{de_ruiter_distinct_2021, mit_medialit_deepfakes}. Talking-head videos expose two channels through which the identity is conveyed, face and voice, and allow an adversary to manipulate either the identity conveyed by a channel or the behavior expressed through it while leaving the other dimension intact. Crossing these dimensions yielded four manipulation families: voice conversion (VC), text-to-speech (TTS), face-swap, and lip-sync. Each manipulation family alters a distinct set of diagnostic cues, allowing us to examine how participants with different levels of auditory access relied on visual, auditory, and cross-modal evidence when judging video authenticity. 

\subsubsection{Dataset}\label{meth:dataset}
We sourced real videos from the High-Definition Talking-Face (HDTF) dataset~\cite{zhang2021flow}, which comprises high-quality, front-facing recordings of over 300 US politicians and newscasters speaking in natural sentences. 

\subsubsection{Standardization}
Following prior work~\cite{zhang2021flow}, we applied a uniform standardization pipeline at two points. Before generation, we center-cropped the speaker's face, resized it to 512$\times$512, resampled the video to 25 fps and the audio to 48 kHz, and normalized the loudness to -$23$ LUFS. These clips served as both real stimuli and as the inputs to generators. After generation, we applied the same parameters to both the real and manipulated outputs and re-encoded every clip as H.264 + AAC-LC MP4. We standardized clips to two lengths, 11 and 23 seconds, to minimize unintended variability in clip quality or presentation.

\subsubsection{Manipulation Grid}
We constructed the stimulus set by independently varying each video's visual and audio components. Videos with authentic audio and visual content formed the real condition, while the remaining combinations produced visual-only, audio-only, or audiovisual manipulations. This let us identify which channel contained the manipulation and compare detection across hearing profiles.

\subsubsection{Generator Selection and Experimental Scope}\label{meth:generators}
We selected one implementation per manipulation family to cover the distinct attacker capabilities defined in Section~\ref{sec:threat_model}. We selected open-source models available at stimulus construction (Q1 2025) that were locally deployable, compatible with our pipeline, and produced plausible outputs after quality filtering. Widely used commercial systems can change weights and models without notice~\cite{Magic_Hour_API}, impose content controls, or require ownership verification~\cite{ElevenLabs_safety}, making stimuli difficult to reproduce. Generator choice also affects ecological validity, since artifacts and operational constraints differ across deployed models. We therefore grounded selection in documented misuse where possible: our face-swap pipeline used InSwapper-128, the model underlying Haotian AI, a real-time package marketed for deepfake fraud scams~\cite{Cross_2026_haotian}. However, such links are rare, as incident reports seldom name the generator; we do not claim these implementations are the most prevalent in practice. 

For visual manipulations, we applied face-swap and lip-sync techniques. For face-swap, we used FaceFusion with InSwapper-128-fp16 and GPEN-BFR-2048~\cite{yang2021gan} to replace the depicted facial identity with a single target photograph, while retaining the original head motion and scene structure. This condition introduced visual inconsistencies associated with facial-identity replacement. For lip-sync, we used SyncTalk~\cite{peng2024synctalk} to modify facial articulation according to a target audio stream while retaining the source speaker's identity, training the model on each target video to generate its manipulated sample. This condition altered visual speech and its consistency with the accompanying audio.

For audio manipulations, we applied text-to-speech (TTS) and voice-conversion (VC) techniques. We used HierSpeech++~\cite{lee2022hierspeech} to synthesize speech from the HDTF transcript. For VC, we used Seed-VC~\cite{liu2024zero} to modify the perceived speaker identity while retaining the linguistic content and timing of the source speech, conditioning the model on the target speaker's voice to generate each converted sample.

\subsubsection{Quality and Piloting}
We generated over 1,000 candidate manipulated clips, and team members manually reviewed each candidate, removing videos with obvious failures, including audible glitches or clipping, incomplete facial replacement, visible blending boundaries, unstable facial features, frame-to-frame flickering, distorted mouth interiors, and perceptible audio-lip desynchronization. We then conducted an internal 10-person pilot test, with reviewers evaluating clips' authenticity and reporting any identified failures. We retained only perceptually plausible clips.

\subsection{Exposure Design}\label{meth:exp_design}
Our in-person study required a trade-off among manipulation coverage, repeated judgments of each stimulus, participant burden, and session duration. We limited the detection task to 30 trials per participant, allowing sessions to be completed in about one hour while retaining sufficient coverage across each manipulation combination. Each participant evaluated 15 authentic and 15 manipulated clips. The manipulated clips comprised 2 audio-only, 4 visual-only, and 9 combined audiovisual clips, as shown in Figure~\ref{fig:study_procedure}.

\subsubsection{Within-participant Constraints}
Each 30-clip set followed the previously described allocation (Figure~\ref{fig:study_procedure}). Within a~set, we required each source-speaker identity and speech utterance to appear only once, limiting familiarity-based shortcuts and preventing participants from directly comparing manipulated and authentic versions of the same person or utterance. For face-swap clips, we also ensured that inserted facial identities were unique within the set. After assigning the clips, we independently randomized their presentation order for each participant to mitigate ordering effects.

\subsubsection{Matrix Construction}
Because each participant viewed 30 of the 300 clips, we distributed clips across sessions while satisfying the within-participant requirements. We constructed 10 unique sets, resulting in at least three hearing-participant judgments per clip. This allowed us to create a stimulus overlap across hearing profiles, and we accounted for residual differences in clip difficulty using a clip-specific random intercept in the statistical analysis. Each DHH participant received a clip set previously assigned to a HP, providing identical stimuli for cross-group comparisons. We selected sets separately for each subgroup so every clip received at least one judgment from each DHH subgroup.

\subsection{Study Procedure}\label{meth:study_procedure}

After providing informed consent, participants completed the pre-study questionnaire covering media consumption, familiarity with deepfakes, deepfake prevalence, and level of concern about synthetic media. They then completed the main detection task using the 30-clip set assigned through the exposure matrix (\S\ref{meth:exp_design}). For each trial, participants viewed one clip and classified it as \textit{real} or \textit{fake}, with replays permitted. After each judgment, they rated their confidence on a 5-point scale and briefly described the cues leading them to their decision. Participants explained their reasoning only after making an authenticity judgment, preventing their speech from overlapping with the stimulus audio and ensuring their responses could be transcribed clearly. After proceeding to the next video, participants could not return to previous trials, preventing retrospective changes to their responses. The session concluded with a post-study questionnaire capturing reflections on the task and a demographic questionnaire. 

We administered the task through a custom web interface, which recorded the assigned stimulus, presentation order, authenticity judgment, confidence rating, replay count, and any changes to the judgment before submission. With participants' consent, we also audio-recorded all sessions and transcribed them verbatim for subsequent analysis.

\subsection{Ethics}~\label{meth:ethics}
Before data collection, we obtained Institutional Review Board (IRB) approval. All participants provided informed consent to participate and to be audio-recorded. We initially compensated participants with $\$25$ for their time. Because recruitment was slower than anticipated, given the specialized nature of the target population, we increased compensation to $\$50$. We discuss further ethical considerations in Section~\ref{sec:ethics}.

\section{Evaluation}\label{sec:evaluation}
We organize the results around our first two research questions. We first report group-level accuracy and identify the performance gap between HPs and DHH participants (\S\ref{eval:results_overall}), addressing~\ref{rq:q1}. We then examine performance by manipulation channel and type, including TTS, VC, lip-sync, and face-swaps (\S\ref{eval:results_manipulation} and~\ref{subsection:results_hcer}), and test whether misplaced confidence compounds these gaps, addressing~\ref{rq:q2}.

\subsection{Overall Detection Performance}\label{eval:results_overall}

Overall accuracy was $80.9\%$ ($[79.3,82.4]$),\footnote{We report 95\% Wilson score confidence intervals as [,].} but differed significantly between HP and DHH participants ($p<.001$). HPs achieved $88.0\%$ accuracy ($[85.7,89.9]$), compared with $76.4\%$ for DHH participants ($[74.2,78.5]$). A binomial GLMM with participant as a random intercept confirmed lower odds of correct classification for DHH participants (OR $=0.41$, $[0.36,0.47]$, $p<.001$). No single DHH subgroup drove this gap. HoH, d/Deaf, and CI participants achieved $73.1\%$--$78.6\%$ accuracy, and all performed significantly below HPs ($p<.001$, GLMM contrasts). d/Deaf participants had the lowest accuracy ($73.1\%$, $[69.1,76.8]$); only the CI versus d/Deaf comparison was significant within DHH groups ($p=.040$).

\subsubsection{Error Decomposition}
We now examine detection errors by type, using the true-positive rate (TPR) to track missed detections and the true-negative rate (TNR) to track clips mistakenly flagged as manipulated. HPs detected manipulated clips and accepted authentic ones at nearly equal rates (TPR $= 87.1\%$ $[83.7, 89.8]$ vs.\ TNR $= 88.8\%$ $[85.6, 91.4]$; $p=.60$), indicating no systematic response bias ($d'=2.35$, $c=0.04$).\footnote{Sensitivity is calculated as $d'=z\mathrm{(TPR)}-z\mathrm{(FPR)}$, with higher values indicating sharper discrimination between real and manipulated clips. Response criterion is calculated as $c=-0.5[z\mathrm{(TPR)}+z\mathrm{(FPR)}]$, with negative $c$ values indicating a bias toward `fake' judgments and positive $c$ values indicating a bias toward `real' judgments.} 

DHH participants classified real clips less accurately than manipulated ones (TPR $= 82.5\%$ $[79.5, 85.0]$ vs.\ TNR $= 70.3\%$ $[66.9, 73.5]$, $p=.0017$), and exhibited reduced sensitivity ($d'=1.47$) along with a non-zero bias ($p=.003$) toward `fake' judgments ($c=-0.20$). A binomial GLMM with group, stimulus type (real vs.\ fake), and their interaction confirmed that this asymmetry differed significantly between HPs and DHH participants (group $\times$ stimulus-type interaction: $p<.001$). Overall, false alarms on real content drove the performance gap rather than missed deepfakes.

\begin{figure}[t!]
 \centering
 \includegraphics[width=0.95\linewidth]{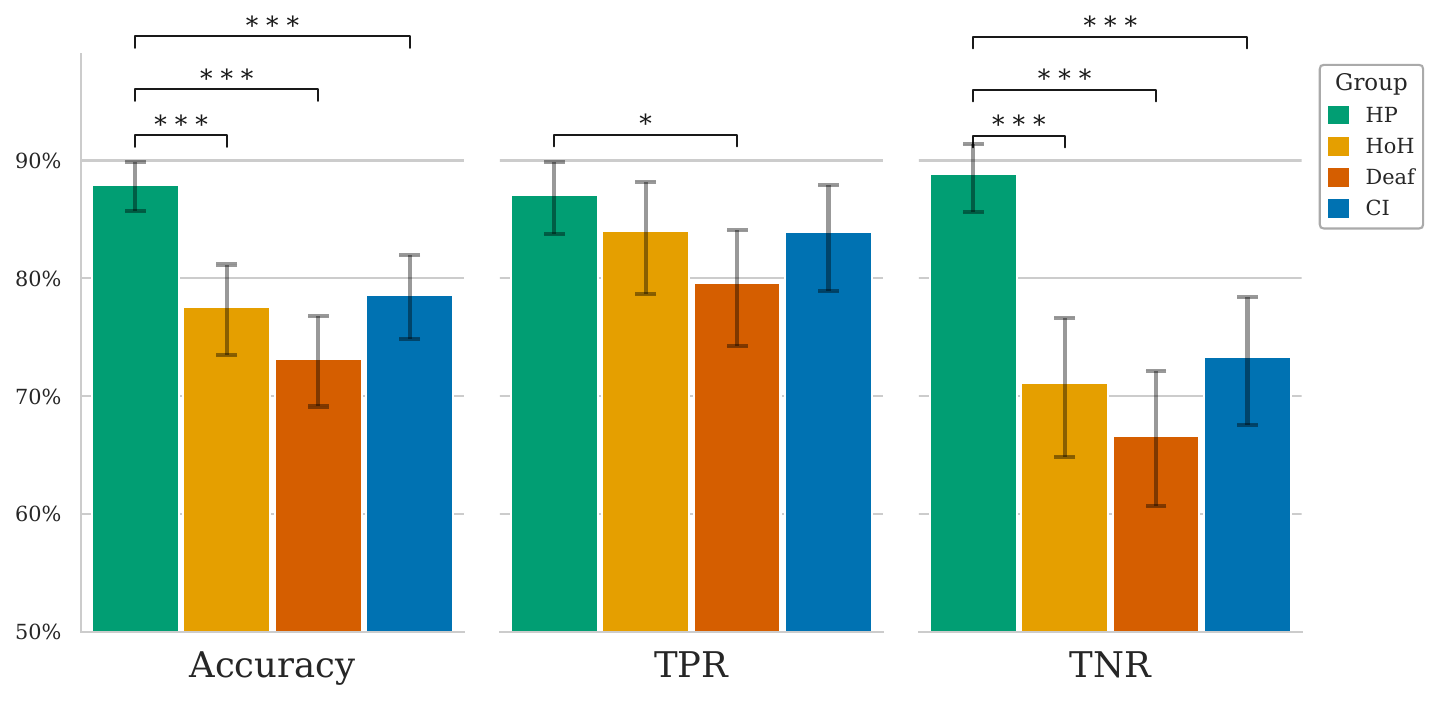}
 \caption{Detection accuracy, true-positive (TPR), and true-negative rates (TNR) by group, with marked Wilson 95\% confidence intervals. Significance markers indicate HP vs.\ DHH subgroups pairwise contrasts (* $p<.05$, *** $p<.001$). }
 \label{fig:acc_tp_tn}
\end{figure}

This TNR gap was present across all three subgroups (Figure~\ref{fig:acc_tp_tn}). Per-group TNRs were $66.7\%$ ($[60.6, 72.3]$) for d/Deaf participants, $71.1\%$ ($[64.8, 76.7]$) for HoH, and $73.3\%$ for CI users ($[67.5, 78.4]$); all were significantly below HPs' TNR of $88.8\%$ (all $p<.001$, GLMM contrasts) but differences among DHH subgroups were not statistically significant (omnibus $\chi^{2}(3)=7.01$, $p=.07$).

HoH and CI users detected manipulations at rates close to HPs (TPR$_\text{HoH}=84.0\%$ $[78.6, 88.2]$, TPR$_\text{CI}=83.9\%$ $[78.9, 87.9]$, vs.\ TPR$_\text{HPs} =87.1\%$), while d/Deaf participants detected somewhat fewer (TPR$_\text{Deaf} =79.6\%$, $[74.2, 84.1]$). Within each DHH subgroup, TPR exceeded TNR by $10.6$-$12.9$ percentage points. This asymmetry was statistically significant for CI users ($p=.005$) but not for HoH and d/Deaf participants ($p=.11$ for both, GLMM contrasts).

These results address~\ref{rq:q1}. HPs detected audiovisual deepfakes more accurately ($88.0\%$) than DHH participants ($76.4\%$, $p <.0001$, subgroups $73\%$--$79\%$), and the gap lay in false alarms on authentic clips rather than missed deepfakes, reflecting both lower perceptual sensitivity ($d'=1.47$ vs.\ $2.35$) and a judgment bias toward `fake' ($c=-.20$ vs.\ $.04$).

\subsubsection{Confidence Calibration}\label{eval:results_confidence}

Performance gaps between groups would be less problematic if incorrect judgments were accompanied by low confidence, but confidence did not align with correctness across groups. HPs were well calibrated (ECE=$0.022$, Brier=$0.097$).\footnote{To assess calibration, we mapped the 5-point scale to the probability range $[0.5,1]$ and computed the Brier score and Expected Calibration Error (ECE) using confidence levels as natural bins. Details are in Appendix~\ref{app:other_mapping}.} In contrast, the aggregate DHH calibration error was roughly four times greater (ECE=$0.100$, Brier=$0.180$), and the miscalibration was concentrated in d/Deaf participants (Deaf: ECE=$0.139$, Brier=$0.201$; HoH: ECE=$0.111$, Brier=$0.180$; CI: ECE=$0.071$, Brier=$0.159$). For DHH participants, confidence was thus a weaker signal of whether a judgment was correct.

\subsection{Performance Across Manipulations}\label{eval:results_manipulation}

The HP--DHH gap varied across manipulations. To address \ref{rq:q2}, we analyzed performance at two levels.\footnote{Tests use binomial GLMMs `outcome $\sim$ predictors + $(1|\mathrm{user\_id}) + (1|\mathrm{vid\_name})$', restricted to the manipulation channel and groups under analysis; see Appendix~\ref{app:glmm_measures}.} We first grouped clips as \textit{audio-only}, \textit{visual-only}, or \textit{audiovisual} to test whether audio-only manipulations drove the gap. We then compared individual deepfake methods (Figure~\ref{fig:study_procedure}) to identify method-specific differences.

\begin{figure}
 \centering
 \includegraphics[width=0.85\linewidth]{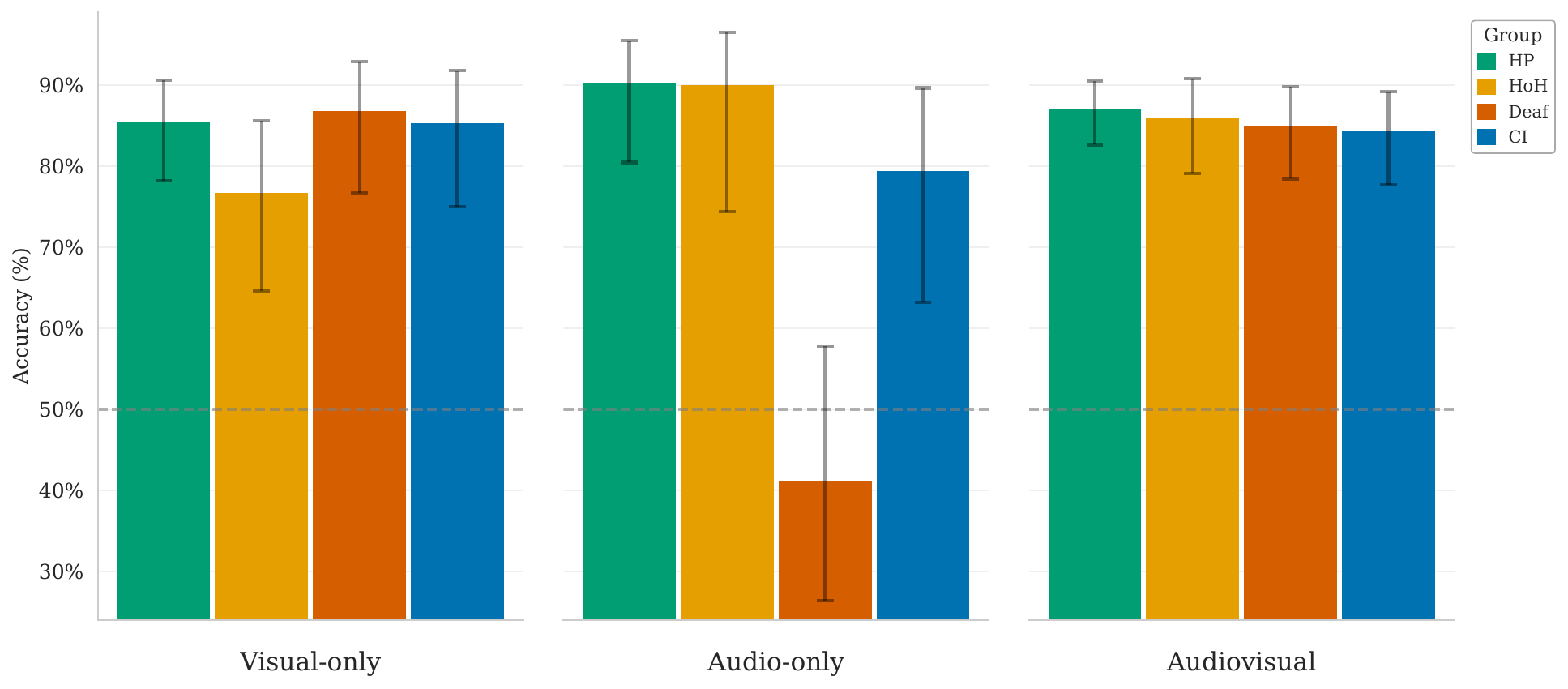}
 \caption{\textbf{Manipulation channel.} Detection accuracy by group across the three manipulation channels with Wilson 95\% confidence intervals. The dashed line marks chance.}
 \label{fig:family_acc}
\end{figure}

\subsubsection{Performance by Manipulation Channel}\label{eval_by_channel}
The HP-DHH gap was largest for audio-only manipulations, where the visual track was real but the audio was altered, as shown in Figure~\ref{fig:family_acc}. HPs detected $90.3\%$ ($[80.5, 95.5]$) of audio-only deepfakes, while HoH detected a similar $90.0\%$ ($[74.4, 96.5]$), and CI detected a lower $79.4\%$ ($[63.2, 89.7]$, $p=.19$ vs.\ HP, GLMM contrasts). d/Deaf participants fared much worse at $41.2\%$ ($[26.4, 57.8]$; $p<.001$ vs.\ HP). They marked $66.7\%$ ($[60.6, 72.3]$) of real clips and $58.8\%$ ($[42.2, 73.6]$) of audio-only fakes as `real' with no significant change in their response ($p=.35$). For d/Deaf viewers, an audio-only manipulation was therefore not a deception but an absence of evidence, where the manipulated channel fell outside the viewer's cue inventory.

For visual-only manipulations, the four groups' detection rates were statistically indistinguishable ($\chi^{2}(3)=2.29$, $p=.51$). HPs $85.5\%$ ($[78.2, 90.6]$), d/Deaf $86.8\%$ ($[76.7, 92.9]$), and CI users $85.3\%$ ($[75.0, 91.8]$) all sat above $85\%$, while HoH descriptively dropped to $76.7\%$ ($[64.6, 85.6]$). That drop was not significant in our sample (HP vs.\ HoH, $p=.182$).

Across audiovisual manipulations with both fake audio and video, all four groups reached similar overall accuracy (HP: $87.1\%$ $[82.6, 90.5]$, HoH: $85.9\%$ $[79.1, 90.8]$, Deaf: $85.0\%$ $[78.5, 89.8]$, CI: $84.3\%$ $[77.7, 89.2]$; $\chi^{2}(3)=.57$, $p=.90$). Given that the d/Deaf group performed much worse on audio-only manipulations, it is notable that they achieved similar accuracy here without access to audio cues.

\begin{figure}[t!]
 \centering
 \includegraphics[width=0.95\linewidth]{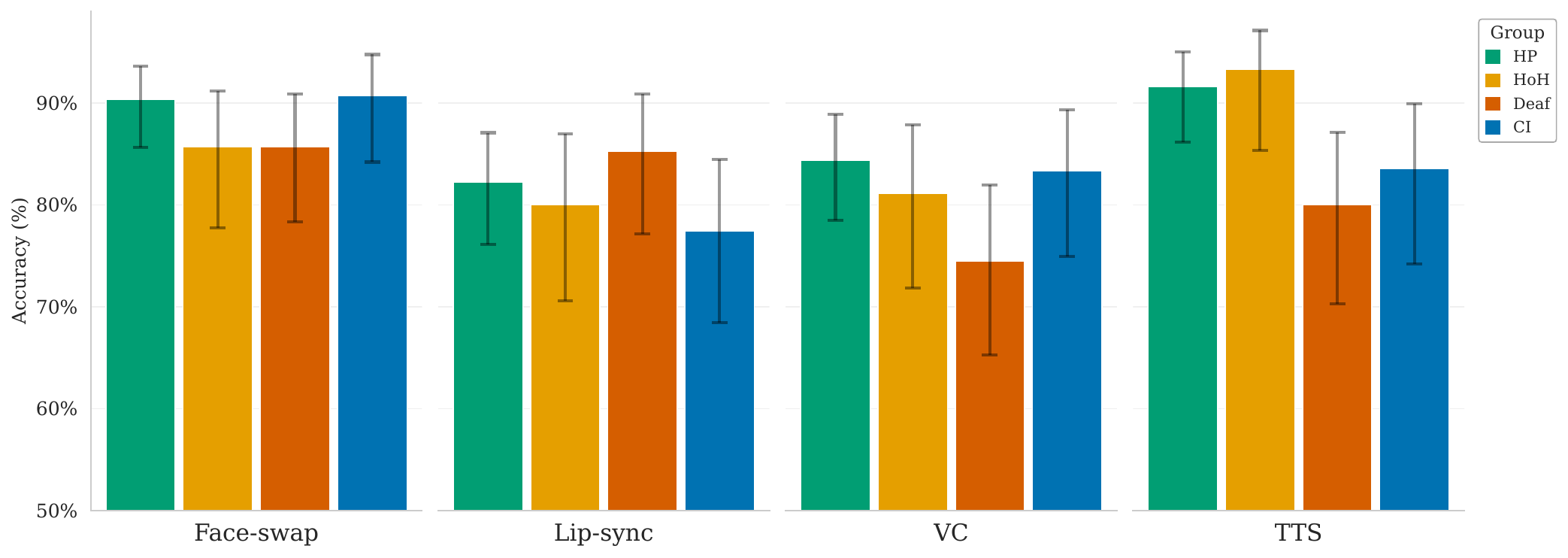}
 \caption{\textbf{Manipulation type.} Detection accuracy by groups across the four manipulation types with Wilson 95\% confidence intervals.}
 \label{fig:manipulation_type_accuracy}
\end{figure}

\subsubsection{Performance by Manipulation Type}
We next examined accuracy by manipulation type (face-swap, lip-sync, VC, and TTS). If a clip contained both visual and audio manipulations, we counted it for both types. Figure~\ref{fig:manipulation_type_accuracy} summarizes the results. Across groups, face-swaps had the highest detection rate ($88.6\%$, $[85.7,91.0]$), followed by TTS ($87.8\%$, $[84.2,90.6]$), with lip-sync and VC both at $81.5\%$ ($[77.7,84.7]$). Face-swaps and TTS were easier to detect than VC ($p=.002$ and $p=.004$), and face-swaps were easier than lip-sync ($p=.001$); other contrasts were not significant.\footnote{Pairwise contrasts are from a GEE refit of the same multi-label technique model with cluster-robust standard errors.}

\paragraph{Audio Manipulations}
For TTS clips, accuracy was the highest for HoH participants  ($93.3\%$, $[85.3, 97.1]$), followed by HP ($91.6\%$, $[86.2, 95.0]$), CI ($83.5\%$, $[74.2, 89.9]$), and d/Deaf ($80.0\%$, $[70.3, 87.1]$). On VC clips, DHH accuracy was lower ($79.6\%$, $[74.6, 83.8]$), with d/Deaf detecting the fewest ($74.5\%$, $[65.3, 82.0]$), compared to HPs ($84.4\%$, $[78.5, 88.9]$, $p=.007$). The cross-group difference was significant for TTS ($\chi^{2}(3)=18.33$, $p<.001$, with HP vs.\ d/Deaf $p=.001$ and HP vs.\ CI $p=.020$) and marginal for VC ($\chi^{2}(3)=7.57$, $p=.056$, with HP vs.\ d/Deaf $p=.007$). Both gaps were concentrated in the cells where only the audio is manipulated (\S\ref{eval_by_channel}).

\paragraph{Visual Manipulations}
On face-swap clips, accuracy was uniformly high across groups ($85.7\%$--$90.8\%$, $\chi^{2}(3)=4.71$, $p=.20$). Although accuracy varied more widely ($77.5\%$ to $85.3\%$) within lip-sync clips, differences across the four groups were not significant ($\chi^{2}(3)=3.45$, $p=.33$). CI users achieved the highest face-swap accuracy ($90.8\%$, $[84.2, 94.8]$) and the lowest lip-sync accuracy ($77.5\%$, $[68.4, 84.5]$), a $13.3$ percentage points within-group gap ($p=.012$). HPs also showed a significant, albeit smaller, gap ($8.1$ percentage points, $p=.045$) in detecting the two manipulation types, but the within-group difference was not significant for d/Deaf or HoH.

If we consider just clips with lip-sync but no audio manipulation, d/Deaf participants showed the highest accuracy ($91.2\%$, $[77.0, 97.0]$, vs.\ HP: $83.9\%$ $[72.8, 91.0]$, HoH: $70.0\%$ $[52.1, 83.3]$, CI: $76.5\%$ $[60.0, 87.6]$), but the differences did not reach statistical significance ($\chi^{2}(3)=4.92$, $p=.18$).

\subsection{High-Confidence Errors}\label{subsection:results_hcer}

\begin{figure}[t!]
 \centering
 \includegraphics[width=0.8\linewidth]{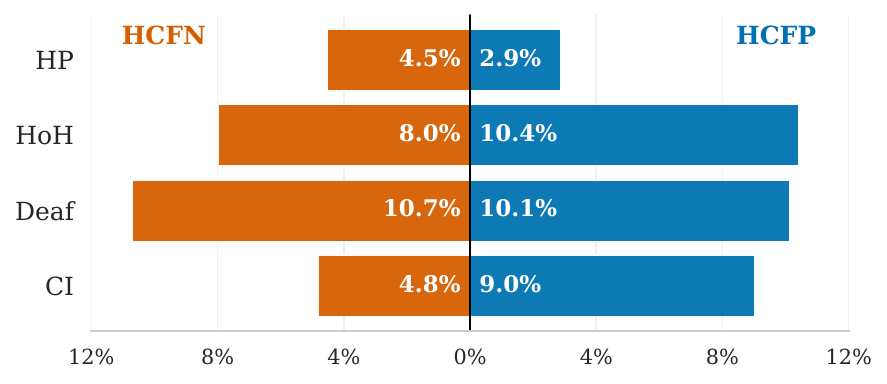}
 \caption{\textbf{High-confidence errors.} Decomposition of high-confidence errors by group into false negatives (HCFN) and false positives (HCFP). Bars show the proportion of high-confidence trials in each error category.}
 \label{fig:hcer}
\end{figure}

We define the High Confidence Error Rate (HCER) as the error rate for judgments with confidence $\geq 4$ out of $5$. High-confidence responses accounted for $71.7\%$ of overall trials, with HPs reporting confidence $\geq 4$ at a higher rate than the DHH cohort ($78.2\%$ $[75.4, 80.7]$ vs.\ $67.6\%$ $[65.2, 70.0]$, $p=.008$).\footnote{HCER cross-group tests use GEE with cluster-robust standard errors clustered on participant. We refit the same model for pairwise contrasts.} Figure~\ref{fig:hcer} shows HCER by group and separates high-confidence errors into false negatives (\textit{HCFN}) and false positives (\textit{HCFP}).

\begin{figure*}[t!]
\centering
\includegraphics[width=0.85\linewidth]{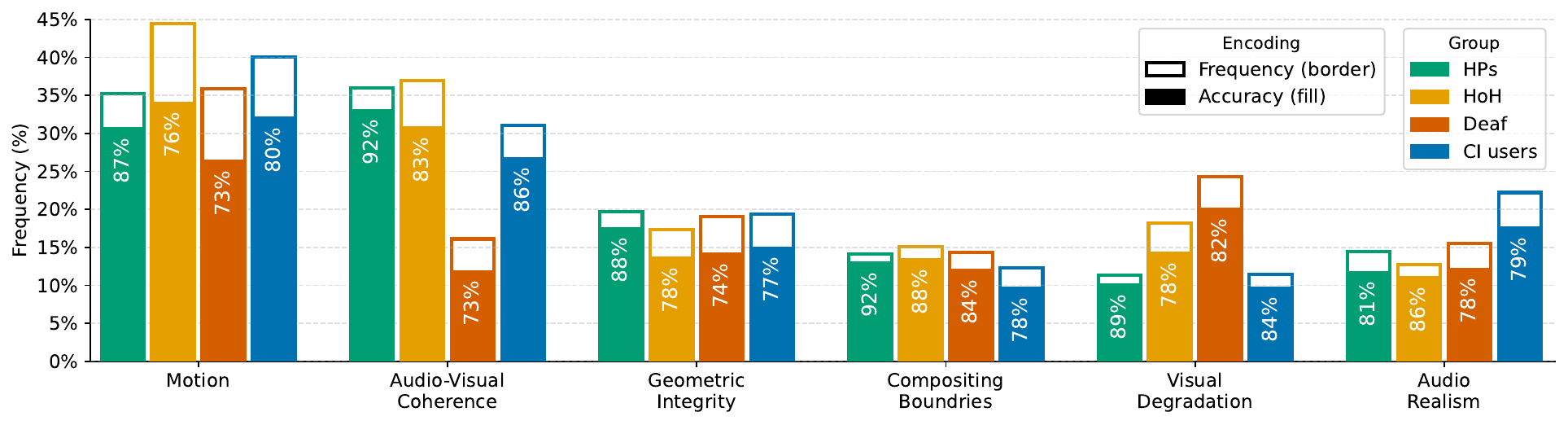}
\caption{Frequency of themes (bar outlines) and their accuracy ($\%$ of filled bars) across participant groups.}
\label{fig:freq_acc}
\end{figure*}

DHH participants made confident errors at more than twice the rate of HPs (HCER $17.7\%$ $[15.5, 20.2]$ vs.\ $7.4\%$ $[5.7, 9.6]$; $p<.001$). The four-group difference was statistically significant ($\chi^{2}(3)=28.29$, $p<.001$), where every DHH subgroup confidently erred more often than HPs (HoH: $18.4\%$, $[14.6, 23.0]$, $p<.001$; Deaf: $20.8\%$, $[16.8, 25.5]$, $p<.001$; CI: $13.9\%$, $[10.6, 18.0]$, $p<.001$). The DHH subgroups did not differ significantly in HCER. Both CI and HoH groups confidently rejected real clips significantly more often than they confidently accepted fakes: CI (HCFP: $9.0\%$ $[6.4, 12.6]$ vs.\ HCFN: $4.8\%$ $[3.0, 7.7]$, $p<.001$) and HoH (HCFP: $10.4\%$ $[7.6, 14.2]$ vs.\ HCFN: $8.0\%$ $[5.5, 11.4]$, $p<.001$). HPs' confident mistakes were more often false negatives, but not significantly so. d/Deaf participants had an elevated risk on both sides (HCFN: $10.7\%$ $[7.8, 14.5]$, HCFP: $10.1\%$ $[7.3, 13.8]$; $p=.12$).

The largest HCFN gap was on audio-only manipulations, where d/Deaf participants confidently misclassified $68.2\%$ ($[47.3, 83.6]$) of fakes as real, compared with $3.6\%$ ($[1.0, 12.3]$) for HPs ($p<.001$). The cross-group HCFN gap disappeared in the visual-only and audiovisual manipulations. This shows that the security risk is driven by modality-specific access to the manipulated channel, not by a uniform deficit across DHH groups.

\section{Thematic Analysis}\label{sec:thematic}
The quantitative results show \textbf{that} detection varied across groups and manipulation types, but \textbf{what} perceptual evidence drives those differences remains unclear (\ref{rq:q3}). In this section, we analyze the cues participants discuss during the detection task, evaluate which reported cues are associated with accurate detection, and test how well between-group differences in cue use explain the accuracy gaps observed in Section~\ref{sec:evaluation}.

\subsection{Coding Methodology}\label{them_meth_coding}

\subsubsection{Corpus and Analysis}\label{them_corp_anal}
We conducted interviews in which participants watched a series of videos, classified each clip as real or fake, and justified each classification (\S\ref{meth:study_procedure}). 
We coded their explanations using an inductive, latent thematic analysis approach~\cite{braun2006using}. 
After initial familiarization, we iteratively developed a cue-label set in which each label was marked as supporting a \textit{real} or \textit{fake} judgment.

Because explanations frequently reference multiple cues, we applied a multi-label scheme and assigned all applicable labels. The codebook contains $419$ distinct labels (268 fake-pointing and 151 real-pointing, a 1.77$\times$ imbalance favoring fake-pointing cues) and yields 7,476 cue occurrences across the corpus (mean of~$2.9$ cues per trial). 

\begin{remarkbox}[colframe=darkblue,colback=lightblue,title=Remark 1]
Participants engaged substantively, with $2.9$ cues and $53.1$ words per video trial, providing rich information. 
\end{remarkbox}\vspace{0.4cm}

We organized the cue codes into a hierarchical structure comprising thirteen themes nested within three perceptual domains (Audio, Visual, and Heuristic; Table~\ref{tab:full_thematic_cues}). Statistical analyses showed high annotator agreement and good saturation for both the overall corpus and within each group; details are provided in Appendix~\ref{app:thematic}.

\subsection{Overall Cue Landscape} \label{them:cue_landscape}
\subsubsection{Cue Output}\label{app:cue_output}
Explanation length was similar across groups, but cue use differed. DHH participants cited more cues per trial than HPs ($3.02$ vs.\ $2.66$, $p<.001$), driven by HoH ($3.68$) and CI users ($3.08$), whereas d/Deaf participants cited the fewest ($2.38$, $p<.001$ below HPs). DHH participants also used a broader vocabulary, $53.6$ distinct cue codes across $30$ trials vs.\ $44.4$ for HPs ($p<.001$), again driven by HoH ($59.7$) and CI users ($57.4$), while d/Deaf participants matched HPs ($44.4$).

\subsubsection{Theme Frequency}\label{them:theme_frequency}

\begin{figure*}[t!]
\centering
\includegraphics[width=0.85\linewidth]{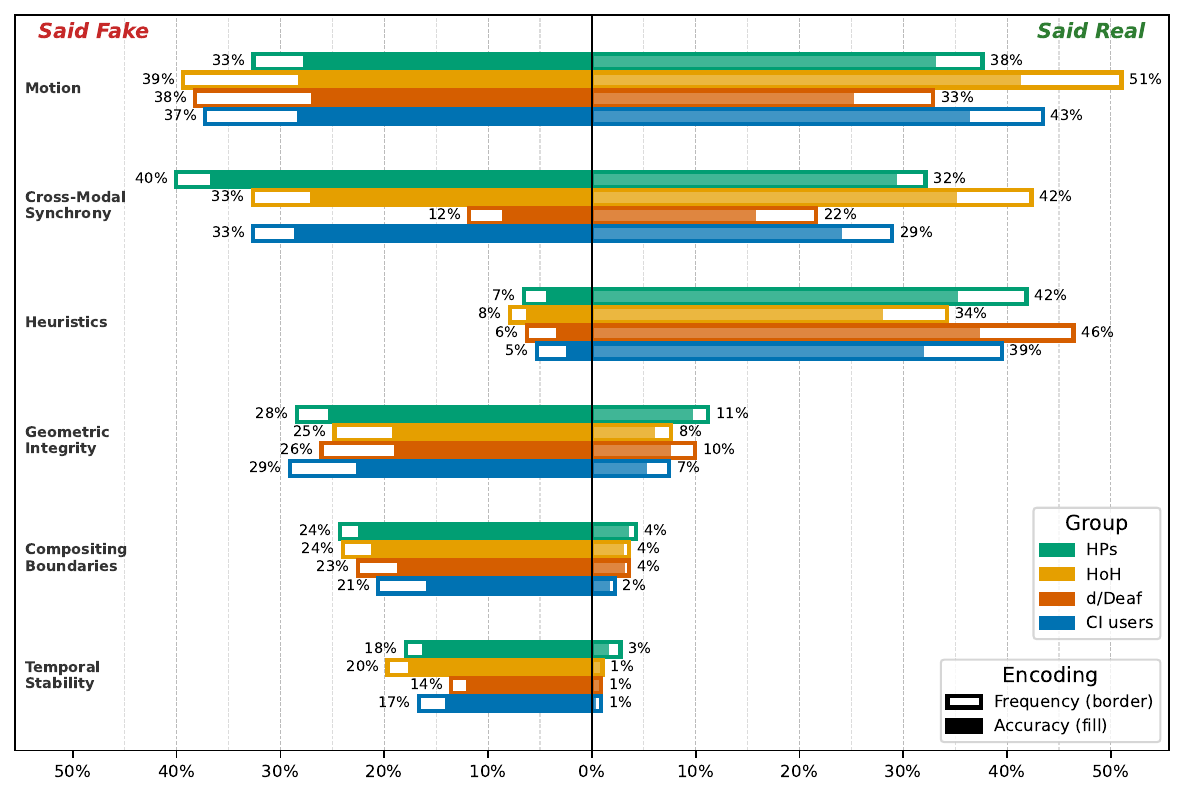}
\caption{\textbf{Fake and Real Detectors.} The left side of this butterfly plot shows each theme's frequency (bar outlines) and accuracy (\% of filled bars) for trials judged as fake, while the right side shows the corresponding values for trials judged as real.}
\label{fig:butterfly}
\end{figure*}

As Figure~\ref{fig:freq_acc} shows, \textit{Motion} ($38.1$\%) and \textit{Cross-Modal Synchrony} ($30.9$\%) dominated. The latter captured perceived audio--visual alignment, such as \textit{``It matches. What I see is what I hear''} (\mbox{HP-23}) and \textit{``prosodic markers are not in alignment with the mouth movement''} (\mbox{Deaf-08}). These two themes were the most cited for the HP, HoH, and CI groups. d/Deaf participants shared \textit{Motion} ($35.9\%$) but not \textit{Cross--Modal Synchrony} ($16.1\%$).\footnote{Due to the sample sizes, most comparisons in this section are not statistically significant.} The second most frequently mentioned cue for d/Deaf participants was \textit{Visual Degradation} ($24.3\%$): image blur, sharpness, and focus. Participants noted cues such as \textit{``it looks very sharp''} (\mbox{Deaf-06}) and \textit{``the glasses were really blurry.''} (\mbox{Deaf-12}). 

A divergence between HP and DHH participants emerged in \textit{Scene Composition}, specifically background-related cues, which DHH participants relied on nearly three times as often as HPs ($17.6\%$ vs.\ $5.9\%$, $p<.001$). When HPs discussed background, they used simple, high-level cues with little elaboration: static (\textit{``[background] is not moving''}, \mbox{HP-10}), virtual (\textit{``a~virtual background at [\ldots] a news station''}, \mbox{HP-15}), or called it \textit{``fake''}. DHH participants considered more factors: person-background integration and outlines (\textit{``It feels like [the depicted person] doesn't belong in this background''}, \mbox{CI-02}), lighting and shadows (\textit{``the brightness in the background does not match the face''}, \mbox{CI-08}), and contextual mismatch (\textit{``a~golf course [in the background while] talking about being responsible and going to a convention so that the context would be a mismatch''}, \mbox{CI-03}).

\begin{remarkbox}[colframe=darkblue,colback=lightblue,title=Remark 2]
DHH participants offered a broader range of cues and gave more detailed rationales, decomposing scenes and integrating observations with prior knowledge.
\end{remarkbox}

\subsection{Theme Uses and Accuracy}\label{them:theme_accuracy}

Themes often functioned as either \emph{fake} or \emph{real detectors}. Figure~\ref{fig:butterfly} shows their frequency and accuracy by judgment. \textit{Motion} and \textit{Cross-Modal Synchrony} were used for both judgments, whereas \textit{Geometric Integrity}, \textit{Compositing Boundaries}, and \textit{Temporal Stability} were used more in fake judgments, and \textit{Heuristics} more in real judgments.

Fake detectors tend to highlight specific anomalies, such as \textit{Compositing Boundaries} and \textit{Temporal Stability}. 
The \textit{Compositing Boundaries} theme captures how regions are blended with their surroundings, including edge quality and mask stability (e.g., \textit{``the face is kind of pasted on to a face''} \mbox{CI-02}), occlusion consistency (\textit{``part of his neck is see-through''} \mbox{HP-06}), and anatomical attachment (\textit{``head just gets detached from his body''} \mbox{(HP-18)}).
Temporal Stability addresses glitches, flicker, and temporal corruptions, invoked by \mbox{CI-11} as \textit{``glitching on the person's chin''}, or by \mbox{HP-17} as \textit{``a black little spot that keeps flickering in and out.''}

When participants used \textit{Compositing Boundaries} and \textit{Temporal Stability}, they reached $97.7\%$ and $97.0\%$ accuracy, respectively, on manipulated clips (freq: $16.8\%$ and $21.8\%$). However, accuracy dropped to $36.1\%$ and $46.6\%$ on real clips (used in only $3.0\%$ and $6.1\%$ of trials). These patterns held for both HP and DHH groups.

Real detectors were exemplified by the \textit{Heuristics} theme, which captured intuitive judgments and familiarity without naming a specific perceptual cue, such as \textit{``I feel something is off''} (\mbox{CI-04}) or \textit{``looks real to me''} (\mbox{HP-26}). They showed the mirror image of the fake-detector themes. When used to justify marking a clip as real, \textit{Heuristics} were highly reliable ($92.7\%$ accuracy). Their performance on manipulated clips (e.g., \textit{``something seemed off,''} \mbox{Deaf-13}) was poor overall, at $39.9\%$, and across all hearing profiles ($29$--$59\%$). 

\textit{Cross--Modal Synchrony} was the only theme that was both accurate and frequently used across both real (acc: $85.9\%$, freq: $29.0\%$) and manipulated clips (acc: $87.1\%$, freq: $32.8\%$). This may be because it can capture both detailed cues -- \textit{``He also has an Adam's apple that's moving as he's talking''} (\mbox{Deaf-10}) -- and slightly more abstract impressions -- 
\textit{``the audio and the lip-sync are not connected''} (\mbox{HoH-12}).

HPs were more accurate than DHH participants when using nearly every theme we developed. The largest difference between HPs and DHH appeared in \textit{Surface Texture} (HP: $89.4\%$ vs.\ DHH: $71.3\%$) -- e.g., \textit{``the skin looks really smooth''} (\mbox{HoH-03}), and \textit{Geometric Integrity} (HP: $88.5\%$ vs.\ DHH: $76.3\%$) -- e.g., \textit{``the nose shape is really weird''} (\mbox{Deaf-03}).  Additionally, \textit{Geometric Integrity} exhibited the largest real-fake disparity, with a $32.6$ percentage-point gap between HP and DHH participants on real clips ($76.7\%$ vs.\ $44.1\%$) and only a $1.5$ percentage-point gap on manipulated clips ($94.3\%$ vs.\ $92.8\%$). DHH participants' attention to fine-grained image details produced more false-positive judgments on real clips. 

\begin{remarkbox}[colframe=darkblue,colback=lightblue,title=Remark 3]
Detailed artifact cues most effectively supported the detection of manipulated clips, but they may have led DHH participants towards false positives. By contrast, participants often relied on less specific intuitions to correctly identify real clips.
\end{remarkbox}

\subsection{Detection by Manipulation Channel}\label{them:manip_channel}
Section~\ref{sec:evaluation} shows that detection accuracy varied systematically across manipulation channels and manipulation types, and that differences across participant groups depended on the manipulated sensory channel.
Participants did not know whether a clip contained a manipulation or which manipulation it contained, so their explanations reflected general inspection strategies rather than manipulation-specific labels. How well those strategies aligned with the manipulated signal helps explain both the similar performance of d/Deaf participants across lip-sync and face-swap manipulations and the gap between these two visual manipulations among CI users.

\subsubsection{\textbf{Audio}}\label{them:manip_audio}
The largest cross-group accuracy gap occurred for audio-only manipulations, where access to the manipulated channel varied most. Participants' explanations corroborated this: even when they invoked audio cues, some participants explicitly qualified what they could hear. \mbox{HoH-05} noted, \textit{``The voice sounds fake. I can't confirm 100\%, because [\ldots] I can't hear it 100\%''}, while \mbox{Deaf-05} explained, \textit{``I can hear their voice. But the words, I can't hear.''}

\paragraph{Audio Realism}\label{them:audio_realism}
This theme encompassed voice quality, naturalness, and audio context (e.g., \textit{``I cannot perceive any kind of emotion of the person's voice''} (\mbox{CI-11}), \textit{``sounds a bit too pitchy''} (\mbox{HoH-10})). It appeared in $14.4\%$ of trials overall, with notable variation across groups (HoH:~$22.2\%$, CI:~$21.0\%$, HP:~$12.7\%$, Deaf:~$4.1\%$), reflecting different verbalization styles. HPs typically mentioned audio only briefly, either flagging or affirming clip authenticity (\textit{``audio is robotic''} \mbox{HP-20}, \textit{``It sounds real''} \mbox{HP-04}). HoH and CI users, by contrast, investigated specific audio properties explicitly, even on ambiguous trials or when not completely certain: \textit{``He raised the pitch of his voice, which [\ldots] seemed quite sophisticated''} (\mbox{CI-01}). When discussed, accuracy on \textit{Audio Realism} tracked each group's overall baseline (HP: $86.4\%$, HoH: $79.0\%$, CI:~$81.3\%$), except d/Deaf participants, whose overall accuracy of $73.1\%$ dropped to $57.6\%$. 

\paragraph{VC vs.\ TTS} 
Across all participants, VC manipulations were more difficult to detect than TTS, and this difference was reflected in participants' cue use. They addressed the \textit{Audio Realism} theme at similar rates across the two audio manipulation types (TTS $11.0\%$, VC $10.6\%$), but its reliability dropped sharply on VC ($62.8\%$ vs.\ $88.6\%$ for TTS). \textit{Cross--Modal Synchrony} was reliable for both TTS ($91.7\%$) and VC ($87.2\%$), but participants cited it more often for TTS ($45.3\%$) than for VC ($29.4\%$). This divergence was driven in part by visually accessible synchrony cues, which appeared more often on TTS trials ($37.8\%$) than on VC trials ($22.5\%$). One likely explanation is that VC more effectively preserved the source speaker's prosody, timing, and speech content, yielding fewer perceptible cross-modal inconsistencies than TTS.

\subsubsection{\textbf{Visual}}\label{them:manip_visual}
Detection accuracy converged across groups when a visual manipulation was present, consistent with shared access to the manipulated channel. However, cue strategies diverged across groups.

For visual manipulations, all groups invoked \textit{Motion} at a~broadly similar rate (HP: $37.9\%$, HoH: $46.7\%$, Deaf: $33.8\%$, CI: $36.8\%$). Their explanations focused on perceived unnaturalness in the speaker's movement, such as  \textit{``the cheeks [\ldots] move in an unusual way''} (\mbox{Deaf-14}). The groups diverged in the specific visual cues they prioritized. 

d/Deaf participants focused more on visible artifacts than participants in other groups. They relied on \textit{Visual Degradation} more than twice as often as the other groups ($36.8\%$ vs.\ $16.7$--$19.1\%$), primarily through the \textit{Blur, Sharpness, and Focus} subtheme ($33.8\%$, acc: $91.3\%$). They also cited \textit{Compositing Boundaries} more frequently ($29.4\%$, acc: $95\%$), indicating greater reliance on image quality and boundary artifacts. \emph{CI~users} instead emphasized whole-face realism, assessing \textit{Geometric Integrity} on $35.3\%$ of trials (acc: $87.5\%$) and \textit{Surface Texture} on $25.0\%$ (acc: $94.1\%$). 

\emph{HoH} participants concentrated instead on eye-related cues, especially \textit{Gaze and Eye Dynamics} ($31.7\%$, compared with $7.4$--$8.9\%$ for the other groups), with $68\%$ accuracy. They also cited \textit{Cross-Modal Synchrony} on $30.0\%$ of trials and \textit{Behavioral and Expression Plausibility} on $15.0\%$, but both themes yielded low accuracy ($68.4\%$ and $55.6\%$). \emph{HPs} showed a balanced distribution across these themes, with no single dominant cue family, except as previously discussed, \textit{Motion}.

\paragraph{Lip-sync} 
d/Deaf participants detected these manipulations most accurately, and their artifact-detection strategy explained why. They described rendering signatures that the lip-sync pipeline actually leaves: mouth blur (\textit{``little blur in his mouth, and it looked fairly obvious to me''}, Deaf-09), throat distortion (\textit{``distortion in their Adam's apple [\ldots] does not look natural''}, Deaf-13), and a disconnect between mouth motion and face stillness (\textit{``it looks like they just put a moving mouth on a still face''}, Deaf-11).

\paragraph{Face-swap} 
These fakes produce artifacts across the entire face, and the cue families used in detection shifted correspondingly. \textit{Compositing Boundaries} were accurate across all groups when invoked ($94$--$100\%$) and were invoked at comparable rates ($26.4\%$ of trials). The advantage seen on face-swaps for CI users may be explained by their greater reliance on themes related to facial identity and whole-face realism, including \textit{Geometric Integrity} (CI: $39.5\%$ vs.\ HP: $35.5\%$, Deaf: $31.1\%$, HoH: $28.6\%$) and \textit{Surface Texture} (CI: $18.5\%$,  HP: $12.0\%$, Deaf: $17.7\%$, HoH: $15.2\%$). This scrutiny was often texture- and surface-focused: \textit{``glitching on the person's chin [\ldots] texture errors''}(\mbox{CI-11}). They also paid greater attention to broader contextual cues, such as lighting ($10.1\%$), which appeared to help more for face-swap manipulations than for lip-sync.

\begin{remarkbox}[colframe=darkblue,colback=lightblue,title=Remark 4]
CI users appeared to have paid more attention to whole-face realism, which helped them detect face-swaps, whereas d/Deaf participants focused more on mouth and throat cues visible in lip-syncs. 
\end{remarkbox}

\subsection{Confidence and Effort}\label{them:confidence}

Although DHH participants reported higher confidence at a given accuracy (\S\ref{eval:results_confidence}), their lower average indicated that their overall confidence was not especially high. We observed that DHH participants were more likely to express a lack of confidence in their initial responses and to seek additional cues before deciding. \mbox{CI-16} said, \textit{``I thought it was real, but something threw me off, and it could be just me. I think it's just too much. I made it fake for that reason.''}. DHH participants made a decision based on an initial set of cues and rapidly incorporated additional cues if their answer seemed insufficient. If they found enough evidence to sway them, they changed their response. On average, HPs made assessments with higher confidence than DHH participants and were less likely to deviate from their initial assessments.

These patterns were reflected in the number of cues used, which was significantly higher for CI users and HoH participants than for HPs (\S~\ref{app:cue_output}). They were also reflected in the depth of some comments, such as those on \textit{Audio Realism} (see Section~\ref{them:audio_realism}).

\begin{remarkbox}[colframe=darkblue,colback=lightblue,title=Remark 5]
DHH participants expressed lower confidence in their decisions and worked harder to identify cues. This effort may have contributed to the higher cue counts and false-positive rates we observed.
\end{remarkbox}

\section{Discussion}\label{sec:discussion}
Our results show that audiovisual deepfake risk depends on the viewer's hearing profile, the channel carrying the manipulation, and the type of manipulation. We discuss the implications for adversarial channel selection, mediated access, and sharing, and propose defense requirements grounded in issues we observed, not as interventions validated by our study.

\subsection{Hearing Profile Creates Distinct Attack Surfaces}\label{disc:attack_surface}

Although d/Deaf and CI participants achieved similar accuracy on visual manipulations, their family-specific results showed different descriptive patterns (\S\ref{eval:results_manipulation}). CI users detected face-swaps significantly more accurately than lip-syncs, whereas d/Deaf participants detected both families at nearly identical rates, with the highest detection rate on lip-sync manipulation across groups. 

The d/Deaf pattern aligns with prior evidence of cross-modal plasticity and multisensory adaptation~\cite{bavelier2002cross, colby2022recognizing}. Lip-sync artifacts concentrate around the mouth, where many d/Deaf viewers obtain visual speech information to assess audio evidence~\cite{grant1999hearing}. Accordingly, d/Deaf participants cited mouth and throat artifacts when evaluating these clips (\S\ref{them:manip_visual}, Remark~4), which may explain why they descriptively detected the best lip-syncs across all groups. For CI users, the face-swap advantage may reflect greater reliance on facial identity cues. CIs prioritize speech intelligibility over the spectral detail needed for speaker recognition~\cite{colby2022recognizing}, and prior work found lower audio-deepfake detection among CI users than HPs~\cite{pasternak2025ci}. CI users may therefore place greater weight on facial than vocal identity cues for speaker recognition~\cite{stevenson2017multisensory}. This could account for the large accuracy gap between face-swaps, which alter identity, and lip-syncs, which preserve it, as well as their frequent use of whole-face cues (\S\ref{them:manip_visual}, Remark~4). However, because the group-by-family interaction was not statistically significant, these within-group contrasts do not establish that manipulation-family effects differ across hearing profiles.

While HoH audio-only accuracy matched HPs, d/Deaf participants classified audio-only fakes and authentic clips as `real' at comparable rates. DHH users should therefore not be treated as a single risk category, as a hearing profile is not a single measure of vulnerability but shapes which evidence is available and which cues receive greater scrutiny. Each hearing profile thus presents a distinct attack surface, defined by the evidence it can inspect and the cues it trusts most.

\subsection{Channel Selection and Evidentiary Absence}\label{disc:attack_channel}

Attack surfaces define adversarial choice in our threat model (\S\ref{sec:threat_model}). The intended message determines which channel must change, while the target's access profile determines where these manipulations matter. Manipulation artifacts outside accessible cues may evade scrutiny, whereas those in closely inspected channels must withstand it. This asymmetry was pronounced for d/Deaf participants, whose detection rate more than doubled when the audio manipulation also included a~visual one ($41.2\%$ vs.\ $85.0\%$). 

While this shows the limits of perceptual verification, it does not, by itself, deceive d/Deaf viewers, as without auditory access, the manipulated message itself fails to reach them. Accessibility layers close this gap, as captions, transcripts, or sign-language interpretation deliver the manipulated message while stripping the acoustic artifacts a listener could use to authenticate it, so the deception succeeds without the manipulation ever facing auditory scrutiny. The same layers can also work as a defense, as we discuss in Section~\ref{disc:accessibility_success_attack}.

Our findings illustrate how differences in access can shape adversarial manipulation selection. Although our empirical results concern hearing profiles, the underlying threat model is not specific to hearing. For any target population, channel-based targeting depends on which evidence the target can access, where manipulation artifacts appear, and how delivery and accessibility layers preserve, remove, or transform that evidence. For blind and low-vision recipients, this relationship may reverse relative to our DHH instantiation: audio may provide accessible authenticity evidence, while visual artifacts remain present but unavailable for inspection~\cite{sharevski2024blind, han2024uncoveringblind}. We did not test this population, so extending the threat model requires population-specific evidence rather than assuming our DHH findings transfer directly.

\subsection{Accessibility Layers Transform Available Evidence}\label{disc:accessibility_success_attack}

Accessibility layers can convey a manipulated message while omitting the audio-specific properties that would expose it. They can also introduce new errors, where inaccurate rendering may distort authentic content or introduce inconsistencies absent from the source, leaving the viewer unable to determine whether an anomaly originates from the source, the manipulation, or the accessibility layer. 

Accessibility layers can also expose otherwise unavailable evidence. Human captioners and interpreters attend to audio as they render it, and captioning practices already convey information beyond linguistic content, including speaker identity, nonspeech sounds, and audio quality~\cite{may2023nonspeech,alonzo2022beyondsubtitles}. 

These practices could extend to authenticity, where caption annotations such as \textit{[distorted voice]} or \textit{[voice does not match speaker]} together with interpreting norms for flagging prosodic anomalies could surface evidence otherwise omitted, complementing system-level verification.

\subsection{Sharing Across Access}\label{disc:translated_self}

Mediated access also shapes what happens after an authenticity judgment. The \textit{``translated deaf self''} captures the ontological insecurity of relying on spoken interpretations one cannot fully verify, creating persistent uncertainty about whether what is said in one's name, and what reaches oneself, is faithful~\cite{young2020translated}. Audiovisual deepfakes extend this insecurity from representation to circulation. A viewer who forwards a~clip without access to its audio evidence may be judged by recipients who can inspect it, treat the share as endorsement, and attribute any resulting error to the sharer's competence or credibility. This risk intensifies across language boundaries, such as when a Deaf ASL user forwards an English-language clip to a hearing, English-speaking network or when an interpreter relays content without being asked to assess its authenticity. Anticipated blame may therefore discourage DHH viewers from sharing content or engaging at all.

\subsection{Excessive Skepticism is a Security Risk}\label{disc:over_skeptcism}

Distrust also extends to excessive skepticism. During the study, DHH participants rejected real clips at a substantially higher rate than HPs, with d/Deaf participants rejecting one in three real clips, consistent with the fake-leaning response bias across all DHH subgroups. Their longer trial times, more frequent replays, and more detailed explanations (\S\ref{app:cue_output} \& Appendix~\ref{app:replay}) indicate that these false alarms reflect deliberate evaluation rather than rushed errors. 

This over-rejection could cost the communication infrastructure DHH users rely on. Systems such as Video Relay Service (VRS), Video Remote Interpreting (VRI), telehealth, and captioned video depend on trust in authentic content as much as on detecting manipulated content. We did not observe such harms directly, but our DHH participants entered the study already estimating a significantly higher prevalence of manipulated media ($p=.002$; Appendix~\ref{app:deepfake_prevalance}), indicating that this distrust predates any single deepfake encounter, and growing awareness of synthetic media may compound it. When false alarms are frequent, people may dismiss authentic messages, doubt clinical instructions, and treat interpreter-mediated communication as suspect. Deepfake defenses for DHH users should therefore aim to reduce both error types, rather than treating missed deepfakes as the only failure.

\subsection{Defenses Must Compensate Perceptual Limitations}\label{disc:defenses}

When perceptual inspection cannot establish authenticity, directing viewers to scrutinize more closely provides no protection. Verification must instead be supported at the system level. Content Credentials can record verifiable assertions about a file's origin and transformation history~\cite{C2PA_specs}. In mediated settings, however, verification cannot stop at the source video. Captioning, interpretation, and relay processing create the representation through which the viewer receives the message, so the system must preserve a verifiable link between that representation and its source. Without it, a viewer may verify the original video's provenance but cannot determine whether the caption or interpretation represents it faithfully. Even linked provenance establishes only the recorded origin and transformations. It does not establish that the message is true or that a derived representation is accurate.

Verification signals must themselves be accessible. An auditory warning remains unavailable to a viewer who cannot hear it, while a visual warning that obscures the speaker's mouth removes a cue d/Deaf participants rely on. Warnings should therefore be delivered across modalities and distinguish verified provenance, detected anomalies, and remaining uncertainty. Binary or low-context warnings may amplify the distrust that DHH users exhibited. Interfaces should explain the basis and uncertainty of each warning and provide an accessible path to independent verification.

Media-literacy interventions should move beyond generator-specific cues and emphasize verification practices that remain useful as deepfake quality improves. These might include verifying the source through previously established contact channels, judging whether the sender and requested action match prior interactions, and seeking corroboration before acting on or resharing a clip. Guidance should also state plainly that a faithful caption establishes what the audio says, not whether the voice is authentic. Because written warnings reach DHH users unevenly across literacy backgrounds~\cite{sharevski_dhh}, interventions should be co-designed with DHH communities and delivered in users' preferred languages and modalities, including ASL.

\label{subsec:limitation}
\subsection{Limitations}\label{disc:limitations}
Our small DHH subgroup samples (15--17 per group) support between-group comparisons but limit the power for contrasts within the DHH group. We therefore treat subgroup-level effects as exploratory. Because we recruited participants locally, our sample may not capture the broader diversity of DHH communities. Future work should replicate these analyses with larger samples drawn from a wider population.

Our stimulus set is correspondingly small, with one generator per family (lip-sync, face-swap, TTS, VC), so observed effects may mix manipulation-type and generator-specific characteristics. Additionally, we were limited to the technology available as of the first quarter of 2025. Other stimuli or generative methods (e.g., face generation, facial manipulation) and newer models might elicit different behavior, warranting further exploration.

Viewing conditions differed from real-world encounters. Participants judged clips on provided equipment, without sender identity, platform context, social signals, or time pressure, and knew they were evaluating deepfakes. In the wild, such contextual cues shape judgments~\cite{ruffin2024does, mink2022deepphish, lovato_diverse_2024}. Results thus reflect controlled inspection, not the full real-world risk profile.

Because participants explained judgments after deciding, their explanations may partly reflect \textit{post-hoc rationalization}. Our goal is not to model what an ideal observer could detect under optimal conditions, but to characterize the cues participants cited. Our main claims rest on between-group comparisons of matched trials, which control for clip difficulty. We report within-group cue correlations for completeness.

We did not include captions, subtitles, or sign-language interpretation in the detection task. We therefore cannot determine whether an accessible rendering adds useful context, removes diagnostic cues, or introduces artifacts mistaken for manipulation. We also did not examine how Deaf signers negotiate authenticity through interpreters or communicate and share uncertainty across linguistic boundaries. Future work should examine how different captioning and interpretation conditions affect detection and how users verify, discuss, and share uncertain content with others.

\section{Conclusion}\label{sec:conclusion}
Susceptibility to audiovisual deepfake deception depends on both the viewer and the manipulation method. We show this through the first empirical comparison of audiovisual deepfake detection across HPs, HoH, d/Deaf, and CI users. In an in-person study with 80 participants, HPs detected audio-only, visual-only, and audiovisual manipulations at high and relatively stable rates. For DHH viewers, however, detection depends on where the manipulation occurs and which perceptual cues it targets. 

These results show that the hearing profile is not a single measure of vulnerability. DHH subgroups exhibit distinct audiovisual attack surfaces, and these differences cannot be inferred from auditory access alone. A lip-sync attack may be more effective against CI users than against d/Deaf viewers, whereas an audio-only attack disproportionately affects d/Deaf viewers. Therefore, prior results based entirely or overwhelmingly on HPs do not generalize to DHH populations. Likewise, a single aggregate accuracy measure can obscure the structure that determines who is at risk, under which manipulation/attack, and why. Audiovisual deepfake threats and defenses should therefore not be evaluated solely against the default user. When defenses treat populations with different sensory access as an afterthought, they may fail the populations most exposed to modality-specific risks. 

\section{Ethical Considerations}\label{sec:ethics}

This study examines how audiovisual deepfake generation techniques affect d/Deaf, hard-of-hearing, and cochlear-implant users, a community underrepresented in security research~\cite{andrew2020review, mack2021accessibility, tran2026toward}. We evaluated the ethical implications when designing the study, assessing effects on participants, and deciding whether to conduct and publish the research, following the Menlo Report.~\cite{menlo}\begin{camready}\footnote{The extended version of ethical considerations is available~\cite{arxiv_version}}\end{camready}

\begin{camready}
\paragraph{Stakeholders}
Study participants, the broader DHH community, the public figures whose likenesses appear in HDTF~\cite{zhang2021flow}, the research team, defenders (detection-tool builders, platforms, accessibility advocates, and policymakers), and adversaries who may use our per-channel results to select manipulations that exploit differences in perceptual access.

\paragraph{Participants}
Our IRB approved the protocol under minimal-risk criteria. Before the study began, participants gave informed consent after being told that the study would present manipulated audiovisual clips and evaluate their perception of deepfakes. All materials were provided in written English with a~verbal or ASL walk-through, and participants could withdraw at any time without penalty. Stimuli were short clips of public political and news speech, screened for graphic, hateful, psychologically distressing, or polarized content. Participants received $\$25$, which we raised to $\$50$ due to insufficient initial interest. Section~\ref{meth:participants} describes the accessibility provisions. 

\paragraph{Data}
Each participant received a random identifier stored separately from the study data. We deleted the original recordings after verifying transcripts and retained only de-identified data on secure institutional servers with access restricted to the research team.

\paragraph{Dual Use and Responsible Disclosure}
Our threat model makes explicit how an attacker could choose among existing manipulation tools based on a target's access profile, and our results show that channel effectiveness varies across hearing profiles, which could support more deliberate attack selection. We therefore report at the granularity of manipulation channel, hearing profile, and cue use, and we do not release per-participant cue-vulnerability profiles or clip-level difficulty rankings. Withholding the findings would not remove the underlying asymmetry. It would leave defenders without the evidence needed to address attackers who already account for access modality. Because attacks that exploit mediated access cannot be countered by asking users to inspect inaccessible evidence, they require system-level defenses. We would reconsider publication if the findings enabled more precise targeting of individuals, gave attackers knowledge they could not reasonably obtain with existing tools, or offered no concrete defensive benefit. Our community collaborators agreed that publication was warranted under these limits.

\paragraph{Positionality}
We are a team of hearing researchers in computer security, HCI, and accessibility. None of us is d/Deaf, hard-of-hearing, or uses a CI, and we do not claim insider status. However, we bring prior experience collaborating with DHH communities on accessibility and security research. Following community usage, we write Deaf for cultural identity, deaf for audiological status, and d/Deaf for both. We describe CI users on their own terms, recognizing that implants are contested within parts of the Deaf community.

\end{camready}

\begin{arxiv}
    
\subsection*{Stakeholders}
We have identified the following primary stakeholder groups potentially impacted by this research:

\begin{itemize}
    \item\textbf{Study Participants.} Hearing and DHH adults who completed our deepfake detection task. 
    \item \textbf{Broader DHH Community.} Deepfake attacks targeting DHH individuals represent a largely understudied threat, and understanding who is vulnerable, under what conditions, and why is an important step towards building defenses that are inclusive by design. Closing this gap in the literature is the motivation for this work.
    \item \textbf{Subjects of deepfakes (HDTF).} Public figures whose likenesses appear in the HDTF dataset~\cite{zhang2021flow} and were used to generate the stimuli.
    \item \textbf{Research Team.}The authors and collaborators who designed and ran the study.
    \item \textbf{Society and defenders.} Detection-tool builders, platforms, accessibility advocates, and policymakers who may use our findings to design more inclusive defenses 
    \item \textbf{Adversaries.} Motivated attackers who may use our threat model and our results per manipulation channel and type to select manipulations that exploit differences in perceptual access across hearing profiles.
\end{itemize}

\subsection*{Ethical Considerations and Impact}

\paragraph*{Respect for Persons} Our study protocol, including the 30 video deepfake trials, was approved by our Institutional Review Board (IRB) under minimal risk criteria. Each participant provided informed consent, in which we outlined the purpose, tasks, data collection, and handling protocols. It also stated that the clips they would view were manipulated and that we were evaluating their perception of deepfakes. We informed participants about the nature of the video content during the consent process. During the study, participants received all consent materials in written English and, in addition, a researcher walked them through all procedures verbally or via ASL. Participants were reminded that they could withdraw from the study at any time without penalty and were compensated for their time.

To ensure the deepfakes presented to participants were nontrivial, the research team reviewed them during quality control. All participating researchers were aware of others' potentially politically divergent views. Each researcher participated voluntarily and could stop at any time without repercussions. The research team also collected data during in-person interviews and transcribed all recordings. None of the sessions or material contained graphic, hateful, or psychologically distressing content. If video content contained polarized views, we removed it during the quality control check to further reduce risk to participants.

\paragraph*{Beneficence} We believe that the risk to individual participants was low: stimuli were short clips of public political and news speech, screened for graphic, hateful, or psychologically distressing content. Additionally, we report results at a granularity that supports defense design (manipulation channel, hearing profile, cue use) without disclosing individually identifying participant cues that could guide targeting. We maximized potential benefits by providing actionable insights to improve deepfake defenses for a population currently absent from detection research and training data. We recognize that a motivated attacker could use these results to target DHH populations through the channel they are least able to audit. That risk falls on the broader DHH community rather than on our participants, and we assess it separately. 

\paragraph*{Justice} 
We recruited and selected participants fairly, including those whose backgrounds and expertise were appropriate for the study goals, while avoiding unnecessary inclusion of additional vulnerable groups (e.g., we included only adult participants). We describe recruitment, eligibility, accessibility provisions (ASL interpreters, ASL-proficient researcher fallback, text-input fallback), and compensation in \S\ref{meth:participants}. Because we conducted all sessions in person, we recruited only participants who could commute to our study location, and we included commuting costs in the participation fee. Additionally, when our initial compensation rate ($\$25$) yielded limited engagement, we increased compensation to $\$50$ to attract more participants. Participants who completed sessions before this change were not retroactively compensated at the higher rate. All participants received compensation for their time, regardless of whether their data were included in the analysis.

\paragraph*{Respect for Law and Public Interest} We complied with our IRB-approved protocol, the HDTF dataset terms of use, and applicable privacy laws. We did not collect, generate, or publish any speech beyond what the dataset's subjects originally recorded.

\paragraph*{Accessibility}
To support accessibility and respect participants' autonomy, we provided ASL interpreters upon request. When the interpreter became unavailable on short notice, we first contacted the participant to reschedule. If rescheduling was not feasible, or if the participant had not requested an interpreter in advance, a member of our research team proficient in ASL conducted the session. As a final fallback, participants could complete the session via text input with a researcher present. After completing the study, participants received monetary compensation for their time.

\paragraph*{Data Management \& De-Identification}
We followed our IRB-approved data management plan and institutional security standards. We assigned each participant a unique, randomized identifier and stored it separately from the study data. After verifying the transcription, we permanently deleted the original audio and video recordings. We retained only de-identified data and stored it on secure, password-protected institutional servers, with access restricted to the research team.

\paragraph*{Responsible Disclosure of Threats}
Our threat model outlines an attacker's selection procedure and discusses which manipulations succeed against each specified hearing profile. After weighing the risks above against the benefit of producing the first systematic evidence on how deepfake detection differs across hearing profiles, which evidence detection-tool builders, platforms, and accessibility advocates currently lack, we concluded that the study should proceed and that results should be published, subject to mitigations described above. Beyond reporting the granularity described above, we do not disclose per-participant cue-vulnerability profiles or clip-level difficulty rankings, so the released material cannot identify which individual stimuli deceived which viewers. We paired each signal identified in our threat model, including accessibility settings, habitual caption use, and disclosed accommodations, with the mitigation intended to reduce the corresponding risk. The DHH community has limited representation in deepfake-detection research, training data, and benchmarks, and may therefore be systematically less well protected by the defenses now being developed.

\paragraph{Dual-Use Assessment}
Publishing the adversary model required us to weigh its defensive value against the capability it may provide. Our threat model provides no new technical capability. Instead, it makes explicit how an attacker could choose among existing manipulation tools based on a~target's access profile. Our empirical results, however, show how channel effectiveness varies across hearing profiles, which could support more deliberate attack selection. This dual-use risk motivates the reporting limits and mitigation above. Withholding our findings would not remove any underlying asymmetry. It would instead leave defenders without the evidence needed to address attackers who already account for access modality. Because many DHH users rely on mediated access to audiovisual content, attacks that exploit these layers cannot be addressed by asking users to exercise greater caution or inspect inaccessible evidence. They require system-level defenses. We therefore publish our findings so defenders can measure and address this disparity. We would reconsider publication if the findings enabled more precise targeting of individuals, gave attackers knowledge they could not reasonably obtain using existing tools, or offered no concrete defensive benefit. Our community collaborators agreed that publication was warranted under these reporting limits.

\paragraph*{Positionality}\label{positional_statement}
We are a team of hearing researchers, working at the intersection of computer security, HCI, and accessibility. We recognize that our backgrounds, identities, and beliefs inevitably shape our research~\cite{positionality
}. None of us is d/Deaf, hard-of-hearing, or uses a cochlear implant (CI), and we do not claim community insider status. Therefore, we cannot fully comprehend the experience of navigating a hearing-default information ecosystem. We do, however, bring substantial prior experience collaborating with d/Deaf and hard-of-hearing communities on accessibility and security research, as well as knowledge of inclusive research practices. We delivered recruitment, consent, instructions, and debriefing in participants' preferred modality: spoken English, written English, or ASL. We treat hearing status as multidimensional. Following community usage, we use Deaf for those who identify with Deaf culture and typically use a signed language, deaf for audiological deafness without presumed cultural affiliation, and d/Deaf when we refer to both. We also recognize that cochlear implants are contested within parts of the Deaf community, and we describe CI users on their own terms rather than as a corrective measure for deafness.
\end{arxiv}

\section*{Acknowledgment}
This material is based on work supported by the National Science Foundation under Grants No. 2055123, 2206950, 2429835, 2446321, and the Omidyar Network Foundation. The authors would like to thank Kaniz Fatima and Andrew Ratayczak for their invaluable assistance with the user interviews.


\bibliographystyle{IEEEtran}
\bibliography{dhh_df}
\newpage
\section*{Appendix}

\appendices
    \providecommand{\qopts}[1]{{\par\vspace{1pt}\small\raggedright\noindent \textit{#1}\par}}

\begin{camready}

\noindent This appendix reports participant demographics, supplementary statistical analyses, and thematic-analysis details. The extended version of this paper~\cite{arxiv_version} also includes the study instruments, additional background and replay analyses, calibration curves, and the full theme table.
\end{camready}
\begin{arxiv}

    \section{Study Protocol \& Instrumentation}\label{app:instrumentation}
    After providing informed consent, participants completed a pre-task questionnaire, the detection task, and a post-task questionnaire. We reproduce each instrument below verbatim. 
    
    \subsection{Pre-Task Questionnaire}\label{app:prequest_questions}
    Before the detection task, participants reported their online media consumption, prior exposure to deepfakes, assumed prevalence of manipulated media, and concern about synthetic media. Participants were asked the following questions:
    \begin{itemize}
        \item Approximately how many hours do you typically spend per day viewing content online? 
        \item Had you ever come across a deepfake online?
        \item If you had to guess, what proportion of online content do you suspect could be deepfakes?
        \item How concerned are you about deepfakes or other computer-generated media?
    \end{itemize}
    
    \subsection{Task and Instructions}
    Participants received the following written instructions before the first trial. The research team, or an ASL interpreter when used, also communicated them to participants.
    
    \begin{tcolorbox}[arc=2pt, fonttitle=\bfseries, colback=white, left=2pt, right=2.5pt, top=2pt, bottom=2pt, title=Study Instructions]
        In this study, you will view short videos. Some may be \textbf{real} (authentic), and others may be \textbf{fake} (altered, manipulated, synthetic, or computer-generated).
        We use ``deepfake'' as an umbrella term for altered or synthetic media.

        \textbf{What You Will Do in This Study:}
        \begin{enumerate}[leftmargin=2em]
            \item \textbf{Watch} each short video carefully. You may replay videos as many times as you need.
            \item \textbf{Decide} if the video is \textbf{real} or \textbf{fake} (digitally manipulated or computer-generated).
            \item \textbf{Rate} how confident you are in your decision.
            \item \textbf{Explain} what cues or reasoning influenced your decision—you can do this by \textbf{speaking} or using \textbf{sign language} with an interpreter.
            \item \textbf{Detailed} and \textbf{specific} explanation is required. General statements (e.g., `looked fake') are not accepted; you must provide details to proceed to the next video.
        \end{enumerate}
        There are \textbf{30 video clips} in total. Please take your time with each one and provide \textbf{detailed reasoning}. We are interested in your perception, not speed.
        
        \textbf{Accessibility \& Captions}
        \begin{itemize}[leftmargin=1.5em]
        \item Captions are turned off during the clips for everyone. We are studying audiovisual judgments—how people combine the available audio with what they see. On-screen text would add extra cues and change what we are measuring.
        \item All instructions and questions are accessible. An ASL interpreter is available upon request.
        \end{itemize}
    \end{tcolorbox}
    
    Participants viewed one clip at a time, classified it as `Real' or `Fake', rated their confidence, and explained the cues that informed the decision. We requested elaboration when a response lacked specificity. Because sessions were audio-recorded, participants provided their explanations only while the clip audio was stopped. Each participant viewed 15 authentic and 15 manipulated clips, yet they were not informed about this ratio in advance. 

    \begin{figure}[!h]
        \centering
        \includegraphics[width=0.8\linewidth]{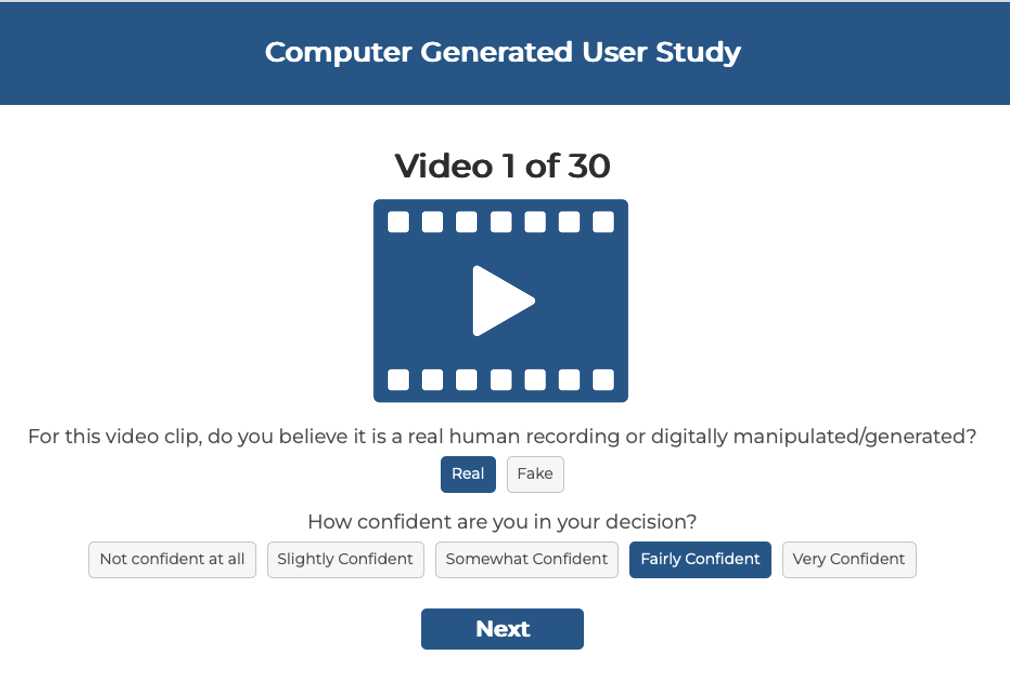}
        \caption{The stimuli as presented to participants, with the instruction, judgment, and confidence responses}
        \label{fig:detection_task}
    \end{figure}

    \subsubsection{Post-task questionnaire}
    After the final trial, participants repeated the concern question, reported their support for deepfake labeling policies and for integrating detection into assistive technologies, and provided demographic information.
    \begin{itemize}
        \item After this study, how concerned are you about deepfakes or other computer-generated media?
        \item What gender do you identify as?
        \item What is your age group? 
        \item Is English your first language? 
        \item What is your highest level of education?
        \item Do you experience any hearing loss or use any assistive hearing devices? 
    \end{itemize}
    
    The following questions appeared only if the participant self-reported hearing loss or use of an assistive hearing device. 
    
    \begin{itemize}
        \item What is your hearing loss (hl) in the \textbf{BETTER} ear? 
        \item Do you use any assistive devices in the \textbf{BETTER} ear?
        \item Do you use any assistive devices in your \textbf{WORSE} ear?
        \item Do you use any assistive technology?  
    \end{itemize}
    
    We combine these responses with participants' communication during the session to assign the d/Deaf, HoH, and CI groups described in Section~\ref{meth:participants}.
    
\end{arxiv}

\section{Demographics and Deepfake Detection}\label{app:participants_characteristics}\label{app:demographic}

\begin{arxiv}
We summarize participant background variables and examine their associations with accuracy, FPR, FNR, confidence, calibration, sensitivity, and criterion. We use Spearman's $\rho$, proportional-odds ordinal regression, or mixed-effects models as appropriate to each outcome's scale and dependence structure. These analyses are observational and aim to contextualize group differences rather than support causal claims.
\end{arxiv}

\begin{camready}
We summarize participant background variables and their association with detection accuracy. These analyses are observational and contextualize group differences rather than support causal claims.
\end{camready}
\begin{arxiv}
\subsection{Demographics}
Assumed prevalence of manipulated media is the only background item that differs between the DHH and HP cohorts, with DHH participants reporting higher estimates. Concern increases descriptively after the task, but no group-level change is statistically significant.
\end{arxiv}
\begin{camready}
Assumed prevalence of manipulated media is the only background item that differs between the DHH and HP cohorts. No background item accounts for the $11.6$ percentage-point accuracy difference between cohorts. Tested against participant-level accuracy within each cohort, no item reaches significance among HPs, and among DHH participants only prior deepfake exposure ($p=.027$) and age ($\rho=-0.31$, $p=.033$) do, so we treat background associations as exploratory.
\end{camready}

\subsubsection{Age, gender, education}
\begin{arxiv}
    Most participants were 18--24 years old ($61.3\%$ HP, $61.2\%$ DHH), followed by 25--34 ($32.3\%$ HP, $16.3\%$ DHH), and 35 or older ($6.5\%$ HP, $22.4\%$ DHH). DHH participants span every age bracket from 35--44 through 65 or older, whereas only two HPs are older than 34. Across the sample, 40 participants identified as male (19 HPs, 21 DHH), 39 as female (11 HPs, 28 DHH), and one as non-binary (HP). The educational level did not distinguish completed degrees from programs in progress, so we report the selected level without treating it as completed attainment. No participant selected an educational level below high school. $19.4\%$ of HPs and $36.7\%$ of DHH participants selected high school, $32.3\%$ and $36.7\%$ selected undergraduate, and $48.4\%$ and $26.5\%$ selected graduate.
\end{arxiv}
\begin{camready}
    Most participants were $18$--$24$ years old ($61.3\%$ of HPs, $61.2\%$ of DHH participants). The DHH cohort was older on average, with $22.4\%$ aged $35$ or older, compared with $6.5\%$ of HPs. Forty participants identified as male ($19$ HPs, $21$ DHH), $39$ as female ($11$, $28$), and one as non-binary (HP). The highest selected education level, which did not distinguish completed degrees from programs in progress, was high school for $19.4\%$ of HPs and $36.7\%$ of DHH participants, undergraduate for $32.3\%$ and $36.7\%$, and graduate for $48.4\%$ and $26.5\%$.
\end{camready}

\subsubsection{Hearing loss, communication, and assistive technology}
\begin{arxiv}
HoH participants predominantly reported moderate hearing loss, d/Deaf participants reported profound bilateral loss (14 of 17), and CI users reported severe-to-profound hearing loss before implantation (16 of 17). All 17 d/Deaf participants used ASL during the session, as did 4 HoH participants and 3 CI users. The remaining DHH participants used spoken language. Hearing aid use was common among HoH participants (10) and rare among d/Deaf participants (3). Among CI users, most had unilateral implants (10), 6 had bilateral implants, and the remaining 1 CI user had a bimodal configuration (CI$+$HA). Every DHH participant selected captions as an assistive technology they use.
\end{arxiv}
\begin{camready}
HoH participants predominantly reported moderate hearing loss, d/Deaf participants reported profound bilateral loss ($14$ of $17$), and CI users reported severe-to-profound loss before implantation ($16$ of $17$). All $17$ d/Deaf participants used ASL during the session, as did $4$ HoH participants and $3$ CI users. The remaining DHH participants used spoken language. Hearing aid use was common among HoH participants ($10$) and rare among d/Deaf participants ($3$). Among CI users, $10$ had unilateral implants, $6$ bilateral, and $1$ a bimodal configuration (CI$+$HA). Every DHH participant selected captions as an assistive technology.
\end{camready}

\begin{arxiv}
\subsubsection{Media consumption.}
Most participants reported viewing online media for 3--4 hours per day (16 HP, 24 DHH), followed by more than 6 hours (4 HP, 9 DHH), 1--2 hours (6 HP, 7 DHH), 5--6 hours (4 HP, 7 DHH), and less than 1 hour (1 HP, 2 DHH), with no significant difference between groups.
\end{arxiv}

\subsubsection{Prior exposure and assumed prevalence}\label{app:deepfake_prevalance}
\begin{camready}
    Most participants reported prior exposure to a deepfake ($77.4\%$ HP, $67.3\%$ DHH), and the cohorts do not differ on this item. They do differ on assumed prevalence (Mann--Whitney $p=.002$, $r=.41$, Holm-adjusted $p=.021$). Among HPs, $71.0\%$ estimated that manipulated content accounts for $6\%$--$30\%$ of online media, whereas $51.0\%$ of DHH participants estimated more than $30\%$ and $10.2\%$ were unsure, an option no HP selected. This higher assumed prevalence aligns with the fake-leaning response tendency reported in Appendix~\ref{app:response_distribution}.
\end{camready}

\begin{arxiv}
    Most participants reported prior exposure to a deepfake ($77.4\%$ HP, $67.3\%$ DHH); an additional $16.1\%$ and $24.5\%$ were unsure, and the cohorts did not differ on this item. Assumed prevalence of deepfakes was the only background item to differ between cohorts (Mann--Whitney $p=.002$, $r=.41$, Holm-adjusted $p=.021$, amongg HPs, $71.0\%$ estimated that manipulated content accounts for $6\%$--$30\%$ of online media and $16.1\%$ estimated more than $30\%$. Among DHH participants, $51.0\%$ estimated more than $30\%$, with $40.8\%$ selecting the $31\%$--$50\%$ range alone, and $10.2\%$ were unsure, an option no HP selected. Mean estimates are similar across DHH subgroups (HoH $3.46$, d/Deaf $3.31$, CI users $3.33$ on the $5$-point scale) and higher than among HPs ($2.58$). The higher assumed prevalence does not translate into detection accuracy. DHH participants detected fewer manipulated clips than HPs yet classified $56.1\%$ of all trials as fake, compared with $49.1\%$ among HPs, under a design in which half of the clips were manipulated (Appendix~\ref{app:response_distribution}).
\end{arxiv}

\begin{arxiv}
\subsubsection{Pre- to post-concern shift}
Participants rated their concern about synthetic media before the first clip and after the final clip on the same five-point item. Mean concern increased from $3.74$ to $3.87$ ($+0.13$ points) among HPs and from $3.88$ to $4.10$ among DHH participants ($+0.22$ points). Mean concern increased by $0.19$ points across all participants, with the largest increase in the HoH subgroup ($+0.4$) and the smallest in HPs ($+0.13$). No within-group change is statistically reliable, and the change does not differ between cohorts. Twenty-two participants (8 HPs, 14 DHH) selected the highest concern level. Overall, 48 participants' concern levels did not change, 21 became more concerned, and 11 became less concerned.
\end{arxiv}

\begin{arxiv}
\subsection{Impact on Detection}\label{app:detect_demographic}

Figure~\ref{fig:demographics} shows detection accuracy across demographic groups, with confidence intervals. To avoid conflating background associations with the $11.6$ percentage-point difference between cohorts, we test each item against participant-level accuracy separately within HPs and DHH participants. No HP item reaches statistical significance ($p<.05$), and within the DHH cohort, prior deepfake exposure ($p=.027$) and age ($\rho=-0.305$, $p=.033$) reach this threshold. We therefore treat the remaining analysis as exploratory.

\begin{figure*}[!ht]
    \centering
    \includegraphics[width=0.95\linewidth]{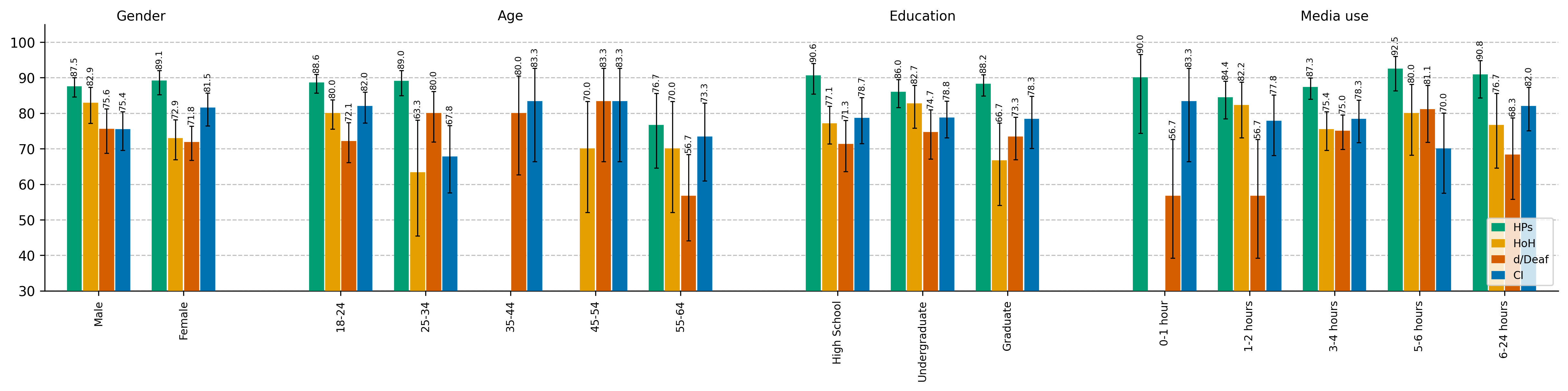}
    \caption{Detection accuracy by hearing profile across demographic and background categories, with $95\%$ confidence intervals.}
    \label{fig:demographics}
\end{figure*}
\end{arxiv}

\section{Supplementary Statistical Analyses}

\subsection{Response-Level Distribution}\label{app:response_distribution}
\begin{arxiv}    
Each participant classified 15 authentic and 15 manipulated clips. Given this balanced exposure, the proportion of `fake' responses provides a descriptive measure of response tendency that complements the criterion values in Section~\ref{sec:evaluation}. 
\end{arxiv}
HPs classified $49.1\%$ of 930 trials as fake, which does not differ from $50\%$ ($p=.784$). DHH participants classified $56.1\%$ of trials as fake, $6.1$ percentage points above the balanced distribution ($p=.003$). DHH response tendency differs from HPs under a participant-clustered GEE ($OR=1.32$, $p=.004$). The three DHH subgroup rates are similar (HoH $56.4\%$, d/Deaf $56.5\%$, CI users $55.3\%$), and we detect no difference among them ($p=.121$). 

\subsection{Clip Judgments}
\label{app:clips_accuracy}

Each of the 300 distinct clips was viewed by at least 3~HPs and at least 1 person within each DHH subgroup, with a~median of 8 views. 
\begin{arxiv}
Median per-clip accuracy was 81.7\% for real clips and 87.5\% for manipulated clips. 
\end{arxiv}
Every viewer correctly classified 82 clips (30 real, 52 manipulated), and 15 videos had accuracy below 50\% (11 real, 4 manipulated).

\begin{arxiv}
We ranked the clips separately by HP and DHH classification accuracy and compared the two rankings. The groups shared more of their easiest clips than their hardest clips. Of the 30 clips with the highest HP accuracy, 17 were also among the 30 clips with the highest DHH accuracy. Of the 30 clips with the lowest HP accuracy, only 4 were also among the 30 clips with the lowest DHH accuracy. Large disagreements between the groups were also lopsided. On 29 clips, HPs classified at least 80\% of responses correctly while DHH participants classified fewer than 60\% correctly (near the 50\% chance level). The reverse pattern, high DHH accuracy with near-chance HP accuracy, occurred on only 7 clips.
\end{arxiv}
\begin{arxiv}
    
\subsubsection{Manipulated Clips}
\label{app:clips_difficulty_manip}
Incorrect judgments varied across manipulation types. Participants classified face-swap most accurately, with 31 of 60 clips correctly classified by each participant (FS$+$TTS: 13/20, FS$+$real: 6/20, FS$+$VC: 12/30), whereas participants classified lip-sync manipulation correctly in 13 of 70 clips (LS$+$TTS: 7/20, LS$+$real: 6/30). The three most difficult manipulation classes with unanimous false acceptance were TTS alone (2/10 unanimous), lip-sync with VC (15/20), and VC alone (5/10). Voice conversion was the only class for which no clip was classified correctly by every participant who viewed it. 

\subsubsection{False Positives on Authentic Clips}
\label{app:false_positive_patterns}
\end{arxiv}
To characterize unsafe distrust, we analyze real clips and how participants classified them. False positives occur across the whole DHH cohort rather than concentrating in a few participants, with a mean of 4.3 of 15 real clips falsely rejected for HoH participants, 4.0 for CI users, and 5.0 for d/Deaf participants, compared to 1.7 for HPs. Nine HPs correctly classified each real clip, compared with one participant in each DHH subgroup. Moreover, only 1 HP rejected more than 5 real clips compared to 45\% of DHH participants. Additionally, real clips that HPs classified as fake did not match those that DHH participants incorrectly classified.

\subsection{Replay Behavior}\label{app:replay}
DHH participants replayed clips more often than HPs, averaging $2.92$ views per trial among HoH participants, $3.00$ among d/Deaf participants, and $3.05$ among CI users, compared with $2.60$ among HPs. They were also less likely to judge a clip before replaying it ($3.9\%$--$10.9\%$ of trials across the three DHH groups, compared with $11.7\%$ for HPs). Accuracy declined with each successive replay count, from $93.2\%$ on trials viewed once to $85.7\%$, $79.2\%$, $72.3\%$, and $64.8\%$ after one, two, three, and four or more replays.

\begin{arxiv}
A participant's mean replay count was only weakly related to their overall accuracy within either cohort (Spearman $\rho=-0.26$ for HPs, $p=.15$; $\rho=-0.13$ for DHH participants, $p=.39$). The pooled correlation of $\rho=-0.29$ therefore reflects the combination of more replaying and lower accuracy among DHH participants. 
The number of replays within a~participant was also correlated with lower accuracy (mean $\rho=-0.20$, $p<.001$), as shown in Figure~\ref{fig:app_replay}. Mean confidence also decreased from $4.71$ on trials viewed once to $3.26$ on trials replayed four or more times. This also correlates with time spent on trial, with DHH participants spending more time per trial overall (HoH: $M = 67.6$s, Deaf: $M = 55.8$s, CI: $M = 58.6$s) than HPs ($M = 39.6$s).

\begin{figure}[!t]
    \centering
    \includegraphics[width=0.9\linewidth]{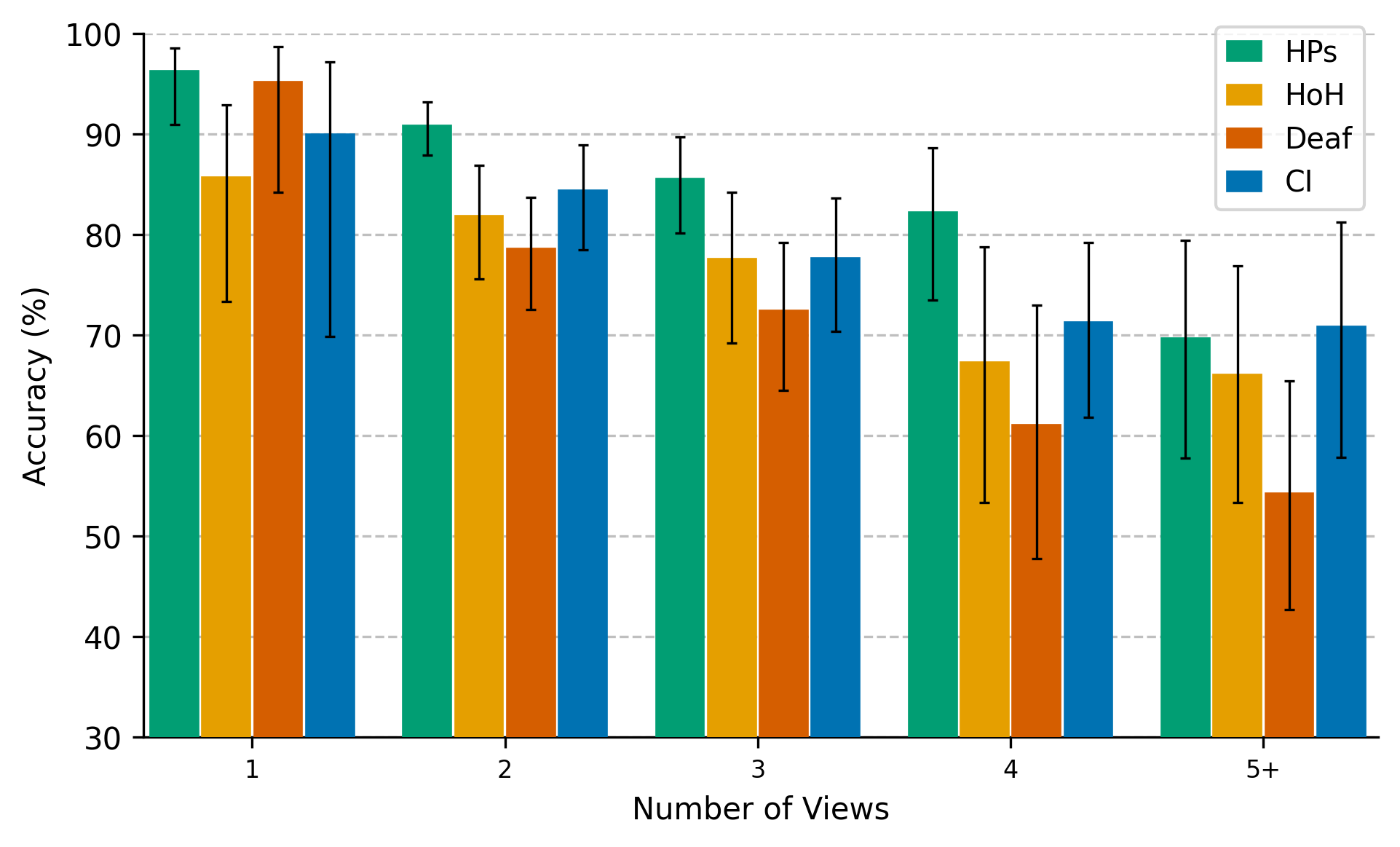}
    \caption{Clip views and classification accuracy across hearing-profile groups. Error bars show $95\%$ confidence intervals.}
    \label{fig:app_replay}
\end{figure}
\end{arxiv}

Replay behavior also differed across hearing profiles for audio-only manipulations, with $1.31$ for HPs, $1.63$ for HoH participants, $1.97$ for CI users, and $2.56$ for d/Deaf participants. HPs replayed audio-only clips less often than authentic clips ($1.31$ vs.\ $1.82$, $p=.003$), whereas d/Deaf participants were the only group to replay them more often. This pattern is consistent with participants repeatedly inspecting accessible visual cues when evidence of manipulation was confined to audio. 
\begin{arxiv}
The channel-specific asymmetry observed in accuracy therefore also appears in replay behavior.
\end{arxiv}

\subsection{Confidence Calibration Mapping}\label{app:other_mapping}

We assess confidence calibration by mapping the $5$-point confidence scale to the probability range $[0.5,1]$. Since participants first make a binary real-or-fake judgment and only then report confidence, the lowest level maps to chance ($0.5$), the highest to $1.0$, and intermediate levels are linearly spaced at $0.625$, $0.75$, and $0.875$. Using these mapped values, we compute the Brier score and the Expected Calibration Error (ECE), treating confidence levels as natural calibration bins.
\begin{arxiv}
Figure~\ref{fig:calibration} shows the relationships between confidence and accuracy.

\begin{figure}[!t]
 \centering \includegraphics[width=0.9\linewidth]{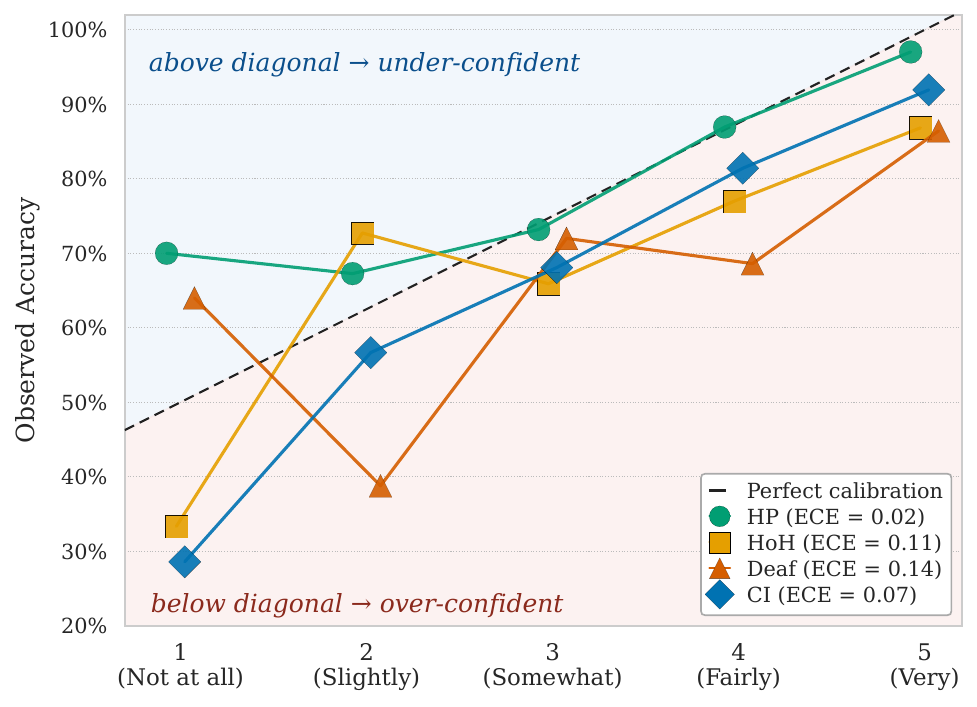}
 \caption{Expected Calibration Error (ECE) by group. The dashed diagonal shows perfect calibration (above: underconfident; below: overconfident). As confidence increased, HPs shifted from underconfident to calibrated, whereas DHH subgroups remained moderately overconfident.}
 \label{fig:calibration}
\end{figure}
\end{arxiv}
\begin{arxiv}
\subsection{GLMM Measures}
\label{app:glmm_measures}

\subsubsection{GLMM Across All Video Types}
\label{app: glmm across video}

A trial-level GLMM shows that channel-level detection patterns differ significantly across participant groups ($\chi^2(9)=45.5$, $p<.001$). HPs' and CI users' performance is relatively stable across manipulation families. At the same time, d/Deaf participants show a significant decline on audio-only clips ($41.2\%$) but high performance on audiovisual clips ($85.0\%$), with the highest performance across all groups on visual-only clips ($86.8\%$) ($\chi^2(3)=39.7$, $p<.001$). HoH participants, by contrast, exhibit the~\textit{weakest} performance on visual-only fakes ($76.7\%$) with a nearly $10$ percentage point gap below the d/Deaf group, despite matching HPs almost exactly on audio-only manipulated clips ($90.0\%$ vs.\ $90.3\%$) ($\chi^2(3)=13.4$, $p=.0038$).

\subsubsection{Channel-specific Performance Differences}
\label{app:channel_spec_diff}
\end{arxiv}
\begin{camready}
    \subsection{GLMM and Channel-specific Performance Differences}
\label{app:glmm_measures}
\end{camready}

We test whether the performance gap between HPs and DHH participants varied across manipulation channels using a trial-level GLMM. The model included correct response as the outcome, hearing group $\times$ manipulation channel as fixed effects, and crossed random intercepts for participant and stimulus.
The significant group $\times$ channel interaction ($\chi^2(6) = 35.08,~p < .001$) confirmed that HP-DHH performance differences are channel-specific. We therefore conducted HP-versus-subgroup simple-effect contrasts within each channel, following protected-testing procedures that do not require additional correction after a significant omnibus interaction.

\section{Thematic-Analysis Details}\label{app:thematic}

\begin{arxiv}

\begin{table*}[!ht]
    \centering
    \caption {\small Theme structure used for thematic analysis of participant-reported audiovisual deepfake detection cues, organized by perceptual domain.}
    \label{tab:full_thematic_cues}.
    \renewcommand{\arraystretch}{1.15}
    \begin{tabular}{p{0.09\textwidth} p{0.21\textwidth} p{0.62\textwidth}}
    \toprule
    \textbf{Domain} & \textbf{Theme} & \textbf{Subthemes} \\
    \midrule
     
    \multirow{2}{*}[-0.6em]{\textsc{Audio}}
      & Audio Realism
      & Audio Quality; Audio Context; Voice Naturalness \\
    \cmidrule(l){2-3}
      &  \multirow{2}{*}[0.1em]{Cross-Modal Synchrony}
      & Lip-sync; Speech-Movement Coordination; Voice-Appearance Consistency; Audio-Visual Scene Coherence \\
     
    \midrule
     
    \multirow{11}{*}[-1.3em]{\textsc{Visual}}
      & \multirow{2}{*}[0.1em]{Scene Composition}
      & Background Realism; Scene Construction and Object Consistency; Depth and Flatness; Person-Background Integration \\
    \cmidrule(l){2-3}
      & Light \& Color Consistency
      & Color and Tone Consistency; Lighting, Shadows, and Reflections \\
    \cmidrule(l){2-3}
      & Surface Texture
      & Skin, Hair, and Surface Texture \\
    \cmidrule(l){2-3}
      & Geometric Integrity
      & Face and Body Geometry \\
    \cmidrule(l){2-3}
      & {Motion}
      & Movement Naturalness; Coordination \& Coupling; Movement Presence \& Absence \\
    \cmidrule(l){2-3}
    & Expressions Plausibility
        & Facial and Body Expression Naturalness; Gaze, Blink, and Eye Dynamics; \\
    \cmidrule(l){2-3}
      & Temporal Stability
      & Temporal Shape Stability \\
    \cmidrule(l){2-3}
      & Compositing Boundaries
      & Anatomical Attachment and Alignment Integrity; Surface Blending and Transition \\
    \cmidrule(l){2-3}
      & \multirow{2}{*}[0.1em]{Visual Degradation}
      & Blur, Sharpness, and Focus; Global Video Quality and Compression; Glitches, Artifacts, and Visual Corruption \\
    \midrule
    \multirow{2}{*}[-0.1em]{\textsc{Heuristic}}
      & Intuition
      & Holistic Impression; Absence-Based Judgment \\
    \cmidrule(l){2-3}
      & Contextual Familiarity
      & Identity Recognition; Contextual Verification \\
     
    \bottomrule   
    \end{tabular}
\end{table*}
\subsection{Developed Themes}
Table~\ref{tab:full_thematic_cues} shows the themes and theme structure.
\end{arxiv}

\subsection{Annotator Agreement}\label{them_annotator_agreement}
To assess the reliability of the multi-label coding, we evaluate agreement in two complementary ways. We compute cue-level agreement over binary item-by-cue assignments using Cohen's $\kappa$ and Gwet's $\gamma_{AC1}$, and per-item overlap between assigned cue sets using Jaccard and Dice. Across the full $(\text{response}, \text{cue})$ matrix, agreement is high ($\gamma_{AC1}=0.997$, $\kappa=0.74$), and remains high once empty-set true negatives (cues neither annotator selected) are excluded. Item-level overlap across $2{,}429$ coded responses is strong, with a mean Dice of $0.73$ and a mean Jaccard of $0.67$.
\begin{arxiv}
Size-weighted values are $0.67$ (Dice) and $0.59$ (Jaccard), and prevalence-weighted estimates are consistent ($\kappa=0.74$, $\gamma_{AC1}=0.989$).
\end{arxiv}

\subsection{Saturation Analysis} \label{them_saturation_anal}
\begin{arxiv}
To assess whether the explanation corpus is sufficiently comprehensive, we perform a cue-saturation analysis of the coded responses. For each axis, we report the smallest sample size $n$ such that, across every subset of size $n$ drawn from the full sample, the union of cues mentioned within the subset covers at least $90$\% of the cues in the pooled corpus. Across all $80$ participants, the average subset coverage reaches $95$\% at $n=31$ (worst-case coverage of $95$\% at $n=48$). Within each participant group, the required sample size for worst-case coverage is approximately two-thirds of the group: $n = 21$ of $31$ for HPs, $n = 9$ of $15$ for HoH, $n = 10$ of $17$ for d/Deaf, and $n = 10$ of $17$ for CI users. The saturation fractions are comparable across groups, despite HPs having nearly twice the participant count of the DHH subgroups, suggesting that the coding vocabulary stabilizes at a similar fraction of the group rather than at a fixed participant count. We therefore interpret within-group cue-frequency contrasts as stable rather than primarily limited by sample size.
\end{arxiv}
\begin{camready}
For each group we report the smallest $n$ such that every subset of $n$ participants covers at least $90\%$ of the cues in the pooled corpus. Worst-case coverage requires roughly two-thirds of each group ($21$ of $31$ HPs, $9$ of $15$ HoH, $10$ of $17$ d/Deaf, and $10$ of $17$ CI users), and across all $80$ participants average coverage reaches $95\%$ at $n=31$. Because the coding vocabulary stabilizes at a similar fraction of each group rather than at a fixed participant count, we treat within-group cue-frequency contrasts as stable rather than limited by sample size.
\end{camready}

\begin{arxiv}
\subsection{Cue Output}\label{app:cue_output2}
Across $2{,}400$ trials, the 80 participants produced $127{,}547$ words and $7{,}476$ cue instances, an average of $53.1$ words per trial ($95$\% Wilson confidence interval $[51.5, 54.8]$) and $2.9$ cues per trial ($[2.7,3.1]$), indicating substantial engagement with the task. Groups did not vary much by word output. Words per trial are statistically indistinguishable between HPs and DHH participants ($51.0$ $[39.9,62.1]$ vs.\ $54.5$ $[46.3,62.7]$, $p=.61$)\footnote{$p$-values in this section are based on GLMM with crossed random intercepts for participant and clip.}, nor for any DHH subgroup against HPs.

Groups differed significantly in the number of cues discussed per trial. DHH participants discussed more cues on average than hearing participants (DHH: $3.02$ $[2.71,3.33]$, vs.\ HP: $2.66$ $[2.43, 2.90]$, $p<.001$). Within DHH subgroups, heterogeneity was considerable. d/Deaf participants cited the fewest cues ($2.38$ $[2.04,2.72]$), significantly below hearing participants ($p < .001$). In contrast, HoH participants cited the most cues of any group ($3.68$ $[3.05,4.31]$, $p<.001$ vs.\ HP), while CI users fell between these extremes ($3.08$ $[2.55,3.62]$, $p<.001$ vs.\ HP).

DHH participants also used a broader range of cues. Across each participant's $30$ trials, DHH participants cited $53.6$ distinct cue codes, compared to $44.4$ for HPs ($p<.001$). HoH ($59.7$; vs.\ HP $p<.001$) and CI users ($57.4$; $p<.001$) again drive the gap, while d/Deaf participants cited the same number of distinct codes as HPs ($44.4$).
\end{arxiv}
\begin{arxiv}
This label asymmetry likely reflects more than the ease of spotting artifacts. Manipulation cues tend to be salient, concrete, and lexically diverse, whereas real judgments often rest on diffuse impressions of naturalness, and the deepfake-detection framing may itself encourage artifact-seeking.

\subsection{Theme Structure}\label{app:full_thematic_cues}
Thematic taxonomy reveals that human deepfake detection involves a mix of technical analysis, cross-modal processing, and intuitive judgment. Visual cues dominate the framework, with $9$ of $13$ themes being visual, suggesting that people rely heavily on visual inconsistencies to detect deepfakes. However, the \textit{``Cross-Modal Synchrony''} theme captures complex cues such as voice-appearance consistency and speech-movement coordination, indicating that people do not analyze audio and video in isolation. The theme decomposition spans from highly technical cues (\textit{``Geometric Integrity''}, \textit{``Temporal Stability''}, \textit{``Compositing Boundaries''}) to purely subjective ones (\textit{``Holistic Impression''} or \textit{``Absence-Based Judgment''}), indicating that both technical and intuitive detection approaches coexist. Contextual knowledge also matters, as the \textit{``Contextual Familiarity''} theme encompasses identity recognition and contextual verification, suggesting that people draw on existing knowledge of public figures or contexts. Participants identify specific technical flaws, with detailed subthemes such as \textit{``Anatomical Attachment and Alignment Integrity''} and \textit{``Surface Blending and Transition.''} Interestingly, poor video quality itself (such as blur, compression, glitches) serves as a detection method, which could be problematic as deepfake quality continues to improve.
\end{arxiv}

\end{document}